\documentclass[dvipsnames]{bioinfo}
\copyrightyear{} \pubyear{}

\usepackage[numbers, sort&compress]{natbib}

\usepackage{etoolbox}
\patchcmd{\thebibliography}{\advance\leftmargin\labelsep}
  {\labelsep=0.7cm \advance\leftmargin\labelsep}{}{}

\usepackage{cleveref} 
\usepackage{float} 
\usepackage{graphicx} 
\usepackage{multirow}
\usepackage{amsmath}
\usepackage{arydshln}
\usepackage[hyphens]{url}
\usepackage{booktabs} 
\usepackage[font=bf,labelsep=period]{caption}
\usepackage{indentfirst}
\usepackage{color}
\usepackage{fancyhdr}
\usepackage{datetime}
\definecolor{mypink}{RGB}{224,8,95}
\definecolor{myblue}{RGB}{31,116,186}
\definecolor{mygreen}{RGB}{35,155,51}
\definecolor{myorange}{RGB}{201,108, 32}
\definecolor{mypurple}{RGB}{122,32,201}
\usepackage{array}
\newcolumntype{P}[1]{>{\centering\arraybackslash}p{#1}}
\newcommand{\subparagraph}{}
\usepackage{titlesec}
\titleformat{\subsection}[block]{\normalfont\normalsize\bfseries}{\thesubsection}{1em}{}
\newcommand{\specialcell}[2][c]{%
	\begin{tabular}[#1]{@{}r@{}}#2\end{tabular}}
\newcommand{\specialcellll}[2][c]{%
	\begin{tabular}[#1]{@{}c@{}}#2\end{tabular}}
\newcommand{\specialcel}[2][t]{%
	\begin{tabular}[#1]{@{}l@{}}#2\end{tabular}}
\usepackage{tabularx}
\newcolumntype{Z}{>{\raggedleft\let\newline\\\arraybackslash\hspace{0pt}}X}
\newcolumntype{Y}{>{\raggedright\let\newline\\\arraybackslash\hspace{0pt}}X}
\newcommand{\etal}{\textit{et al}. }
\newcommand{\ie}{\textit{i}.\textit{e}., }
\newcommand{\eg}{\textit{e}.\textit{g}., }
\usepackage{url}

\usepackage[flushleft]{threeparttable}
\usepackage{ragged2e}

\access{}
\appnotes{}

\begin{document}
\firstpage{1}

\subtitle{}

\title{Nanopore Sequencing Technology and Tools for Genome Assembly: Computational Analysis of the Current State, Bottlenecks and Future Directions}
\author[\textit{Senol Cali et~al}.]{Damla Senol Cali\,$^{\text{\sfb 1,}*}$, Jeremie S. Kim\,$^{\text{\sfb 1,3}}$, Saugata Ghose\,$^{\text{\sfb 1}}$, Can Alkan\,$^{\text{\sfb 2}*}$ \\and Onur Mutlu\,$^{\text{\sfb 3,1}*}$}
\address{$^{\text{\sf 1}}$Department of Electrical and Computer Engineering, Carnegie Mellon University, Pittsburgh, PA, USA\\
$^{\text{\sf 2}}$Department of Computer Engineering, Bilkent University, Bilkent, Ankara,Turkey\\
$^{\text{\sf 3}}$Department of Computer Science, Systems Group, ETH Z\"urich, Z\"urich, Switzerland}

\corresp{$^\ast$To whom correspondence should be addressed.}

\history{}

\editor{}
	
\abstract{Nanopore sequencing technology has the potential to render other sequencing technologies obsolete with its ability to generate long reads and provide portability. However, high error rates of the technology pose a challenge while generating accurate genome assemblies. The tools used for nanopore sequence analysis are of critical importance as they should overcome the high error rates of the technology. Our goal in this work is to comprehensively analyze current publicly available tools for nanopore sequence analysis to understand their advantages, disadvantages, and performance bottlenecks. It is important to understand where the current tools do not perform well to develop better tools. To this end, we 1) analyze the multiple steps and the associated tools in the genome assembly pipeline using nanopore sequence data, and 2) provide guidelines for determining the appropriate tools for each step. Based on our analyses, we make four key observations: 1) The choice of the tool for basecalling plays a critical role in overcoming the high error rates of nanopore sequencing technology. 2) Read-to-read overlap finding tools, GraphMap and Minimap, perform similarly in terms of accuracy. However, Minimap has a lower memory usage and it is faster than GraphMap. 3) There is a trade-off between accuracy and performance when deciding on the appropriate tool for the assembly step. The fast but less accurate assembler Miniasm can be used for quick initial assembly, and further polishing can be applied on top of it to increase the accuracy, which leads to faster overall assembly. 4) The state-of-the-art polishing tool, Racon, generates high-quality consensus sequences while providing a significant speedup over another polishing tool, Nanopolish. We analyze various combinations of different tools and expose the tradeoffs between accuracy, performance, memory usage and scalability. We conclude that our observations can guide researchers and practitioners in making conscious and effective choices for each step of the genome assembly pipeline using nanopore sequence data. Also, with the help of bottlenecks we have found, developers can improve the current tools or build new ones that are both accurate and fast, in order to overcome the high error rates of the nanopore sequencing technology.\\[10pt]
}

\maketitle

\vspace{-15pt}
\section{Introduction} \label{sec:introduction}
Next-generation sequencing (NGS) technologies have revolutionized and dominated the genome sequencing market since 2005, due to their ability to generate massive amounts of data at a faster speed and lower cost \cite{van2014ten, xin2013accelerating, shendure2017dna}. The existence of successful computational tools that can process and analyze such large amounts of data quickly and accurately is critically important to take advantage of NGS technologies in science, medicine and technology.\par
Since the whole genome of most organisms cannot be sequenced all at once, the genome is broken into smaller fragments. After each fragment is sequenced, small pieces of DNA sequences (\ie reads) are generated. These reads can then be analyzed following two different approaches: read mapping and \emph{de novo} assembly. Read mapping is the process of aligning the reads against the reference genome to detect variations in the sequenced genome. \emph{De novo} assembly  is the method of combining the reads to construct the original sequence when a reference genome does not exist~\cite{steinberg2017building}. Due to the repetitive regions in the genome, the short-read length of the most dominant NGS technologies (\eg100-150 bp reads) causes errors and ambiguities for read mapping \cite{treangen2011repetitive,firtina2016genomic}, and poses computational challenges and accuracy problems to \emph{de novo} assembly \cite{alkan2011limitations}. Repetitive sequences are usually longer than the length of a short read and an entire repetitive sequence cannot be spanned by a single short read. Thus, short reads lead to highly-fragmented, incomplete assemblies \cite{lu2016oxford,alkan2011limitations,magi2017nanopore}. However, a long read can span an entire repetitive sequence and enable continuous and complete assemblies. The demand for sequencing technologies that can produce longer reads has resulted in the emergence of even newer alternative sequencing technologies.\par
Nanopore sequencing technology \cite{clarke2009continuous} is one example of such technologies that can produce long read lengths. Nanopore sequencing is an emerging and a promising single-molecule DNA sequencing technology, which exhibits many attractive qualities, and in time, it could potentially surpass current sequencing technologies. Nanopore sequencing promises high sequencing throughput, low cost, and longer read length, and it does \emph{not} require an amplification step before the sequencing process \cite{marx2015nanopores, branton2008potential, laver2015assessing, ip2015minion}.\par
Using biological nanopores for DNA sequencing was first proposed in the 1990s \cite{kasianowicz1996characterization}, but the first nanopore sequencing device, MinION \cite{minionwebpage}, was only recently (in May 2014) made commercially available by Oxford Nanopore Technologies (ONT). MinION is an inexpensive, pocket-sized, portable, high-throughput sequencing apparatus that produces data in real-time. These properties enable new potential applications of genome sequencing, such as rapid surveillance of Ebola, Zika or other epidemics \cite{quick2016real}, near-patient testing \cite{quick2014reference}, and other applications that require real-time data analysis. In addition, the MinION technology has two major advantages. First, it is capable of generating ultra-long reads (\eg 882 kilobase pairs or longer \cite{jain2017nanopore,lomanultralong}). MinION's long reads greatly simplify the genome assembly process by decreasing the computational requirements \cite{lu2016oxford,madoui2015genome}. Second, it is small and portable. MinION is named as the first DNA sequencing device used in outer space to help the detection of life elsewhere in the universe with the help of its size and portability \cite{nasawebpage}. With the help of continuous updates to the MinION device and the nanopore chemistry, the first nanopore human reference genome was generated by using only MinION devices \cite{jain2017nanopore}.\par
Nanopores are suitable for sequencing because they:
\vspace{-5pt}
\begin{itemize}
\item Do not require any labeling of the DNA or nucleotide for detection during sequencing,
\item Rely on the electronic or chemical structure of the different nucleotides for identification,
\item Allow sequencing very long reads, and
\item Provide portability, low cost, and high throughput.
\end{itemize}
\vspace{-5pt}
\par
Despite all these advantageous characteristics, nanopore sequencing has one major drawback: high error rates. In May 2016, ONT released a new version of MinION with a new nanopore chemistry called R9 \cite{r9webpage}, to provide higher accuracy and higher speed, which replaced the previous version R7. Although the R9 chemistry improves the data accuracy, the improvements are not enough for cutting-edge applications. Thus, nanopore sequence analysis tools have a critical role to overcome high error rates and to take better advantage of the technology. Also, \emph{faster} tools are critically needed to 1) take better advantage of the real-time data production capability of MinION and 2) enable real-time data analysis.\par
Our goal in this work is to comprehensively analyze current publicly-available tools for nanopore sequence analysis\footnote{We note that our manuscript presents a checkpoint of the state-of-the-art tools at the time the manuscript was submitted. This is a fast moving field, but we hope that our analysis is useful, and we expect that the fundamental conclusions and recommendations we make are independent of the exact versions of the tools.} to understand their advantages, disadvantages, and bottlenecks. It is important to understand where the current tools do \emph{not} perform well, to develop better tools. To this end, we analyze the tools associated with the multiple steps in the genome assembly pipeline using nanopore sequence data in terms of accuracy, speed, memory efficiency, and scalability.\par

\vspace{-19pt}
\section{Genome Assembly Pipeline Using Nanopore Sequence Data} \label{sec:steps-and-tools}
\begin{figure}[H]
\centering
\includegraphics[width=\columnwidth,height=\textheight,keepaspectratio]{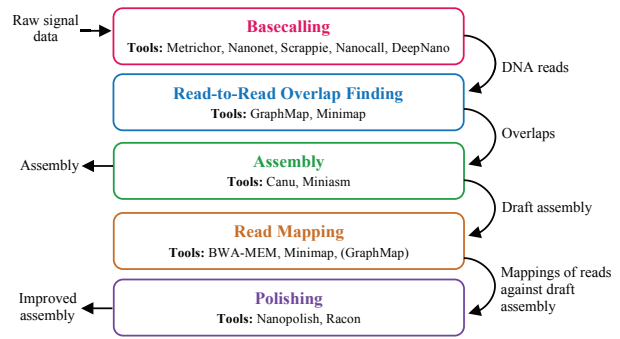}
\caption{The analyzed genome assembly pipeline using nanopore sequence data, with its five steps and the associated tools for each step.}\label{fig:pipeline}
\vspace{-20pt}
\end{figure}

We evaluate the genome assembly pipeline using nanopore sequence data. Figure~\ref{fig:pipeline} shows each step of the pipeline and lists the associated existing tools for each step that we analyze.\par
The output of MinION is raw signal data that represents changes in electric current when a DNA strand passes through nanopore. Thus, the pipeline starts with the raw signal data. The first step, \emph{basecalling}, translates this raw signal output of MinION into bases (A, C, G, T) to generate DNA reads. The second step computes all pairwise read alignments or suffix-prefix matches between each pair of reads, called \emph{read-to-read overlaps}. Overlap-layout-consensus (OLC) algorithms are used for the assembly of nanopore sequencing reads since OLC-algorithms perform better with longer error-prone reads \cite{pop2009genome}. OLC-based assembly algorithms generate an \emph{overlap graph}, where each node denotes a read and each edge represents the suffix-prefix match between the corresponding two nodes. The third pipeline step, \emph{genome assembly}, traverses this overlap graph, producing the layout of the reads and then constructing a draft assembly. To increase the accuracy of the assembly, further \emph{polishing}, \ie post-assembly error correction, may be required. The fourth step of the pipeline is mapping the original basecalled reads to the generated draft assembly from the previous step (\ie read mapping). The fifth and final step of the pipeline is polishing the assembly with the help of mappings from the previous step.\par
We next introduce the state-of-the-art tools used for each step.
\vspace{-10pt}
\subsection{Basecalling}\label{sec:basecalling}
When a strand of DNA passes through the nanopore (which is called the \emph{translocation} of the strand through the nanopore), it causes drops in the electric current passing between the walls of the pore. The amount of change in the current depends on the type of base passing through the pore. Basecalling, the initial step of the entire pipeline, translates the raw signal output of the nanopore sequencer into bases (A, C, G, T) to generate DNA reads. Most of the current basecallers divide the raw current signal into discrete blocks, which are called \emph{events}. After event-detection, each event is decoded into a most-likely set of bases. In the ideal case, each consecutive event should differ by one base. However, in practice, this is not the case because of the non-stable speed of the translocation. Also, determining the correct length of the homopolymers (\ie repeating stretches of one kind of base, \eg AAAAAAA) is challenging. Both of these problems make \emph{deletions} the dominant error of nanopore sequencing \cite{clivebrownwebpage,de2017sequencer}. Thus, basecalling is the most important step of the pipeline that plays a critical role in decreasing the error rate.\par
We analyze five state-of-the-art basecalling tools in this paper (Table~\ref{table:step1}). For a detailed comparison of these and other basecallers (including Albacore \cite{albacorenews}, which is not freely available,  and Chiron \cite{teng2017chiron}), we refer the reader to an ongoing basecaller comparison study \cite{basecallercomparisongithub}. Note that this ongoing study does \emph{not} capture the accuracy and performance of the entire genome assembly pipeline using nanopore sequence data.
\vspace{-10pt}
\subsubsection*{Metrichor}
Metrichor \cite{metrichorwebpage} is ONT's cloud-based basecaller, and its source code is not publicly available. Before the R9 update, Metrichor was using Hidden Markov Models (HMM) \cite{eddy1996hidden} for basecalling \cite{r9webpage}. After the R9 update, it started using recurrent neural networks (RNN) \cite{schuster1997bidirectional, pearlmutter2008learning} for basecalling \cite{r9webpage}.
\vspace{-5pt}
\subsubsection*{Nanonet}
Nanonet \cite{nanonetwebpage} has also been developed by ONT, and it is available on Github \cite{nanonetgithub}. Since Metrichor requires an Internet connection and its source code is not available, Nanonet is an offline and open-source alternative for Metrichor. Nanonet is implemented in Python. It also uses RNN for basecalling \cite{nanonetwebpage}. The tool supports multi-threading by sharing the computation needed to call each single read between concurrent threads. In other words, only one read is called at a time.
\vspace{-5pt}
\subsubsection*{Scrappie}
Scrappie \cite{scrappiewebpage} is the newest proprietary basecaller developed by ONT. It is named as the first basecaller that explicitly addresses basecalling errors in homopolymer regions. In order to determine the correct length of homopolymers, Scrappie performs transducer-based basecalling \cite{clivebrownwebpage}. For versions R9.4 and R9.5, Scrappie can perform basecalling with the raw current signal, without requiring event detection. It is a C-based local basecaller and is still under development \cite{clivebrownwebpage}.
\vspace{-5pt}
\subsubsection*{Nanocall}
Nanocall \cite{david2016nanocall} uses Hidden Markov Models for basecalling, and it is independently developed by a research group. It was released before the R9 update when Metrichor was also using an HMM-based approach for basecalling, to provide the first offline and open-source alternative for Metrichor. However, after the R9 update, when Metrichor started to perform basecalling with a more powerful RNN-based approach, Nanocall's accuracy fell short of Metrichor's accuracy \cite{nanocallgithub}. Thus, although Nanocall supports R9 and upper versions of nanopore data, its usefulness is limited \cite{nanocallgithub}. Nanocall is a C++-based command-line tool. It supports multi-threading where each thread performs basecalling for \emph{different} groups of raw reads.
\vspace{-5pt}
\subsubsection*{DeepNano}
DeepNano \cite{bovza2017deepnano} is also independently developed by a research group before the R9 update. It uses an RNN-based approach to perform basecalling. Thus, it is considered to be the first RNN-based basecaller. DeepNano is implemented in Python. It does not have multi-threading support.
\begin{table*}[b]
\caption{State-of-the-art nanopore basecalling tools.}\label{table:step1}
\begin{tabularx}{\textwidth}{p{1.2cm} p{3.3cm} p{6.0cm} X Z}
\hline
Tool & Strategy & Multi-threading Support & Source & Reference\\ 
\hline
Metrichor & Recurrent Neural Network & (cloud-based) & https://metrichor.com/ & \cite{metrichorwebpage}\\
Nanonet & Recurrent Neural Network & with \texttt{-jobs} parameter & https://github.com/nanoporetech/nanonet & \cite{nanonetwebpage}\\
Scrappie & Recurrent Neural Network & with \texttt{export OMP\_NUM\_THREADS} command & https://github.com/nanoporetech/scrappie & \cite{scrappiewebpage}\\
Nanocall & Hidden Markov Model & with \texttt{--threads} parameter & https://github.com/mateidavid/nanocall & \cite{david2016nanocall}\\
DeepNano & Recurrent Neural Network & no support; split dataset and run it in parallel & https://bitbucket.org/vboza/deepnano & \cite{bovza2017deepnano}\\
\hline
\end{tabularx}
\vspace{-15pt}
\end{table*}
\begin{table*}[t]
\vspace{-5pt}
\caption{State-of-the-art read-to-read overlap finding tools.}\label{table:step2}
\begin{tabularx}{\textwidth}{p{1.2cm} p{3.3cm} p{6.0cm} X Z}
\hline
Tool & Strategy & Multi-threading Support & Source & Reference\\ 
\hline
GraphMap & \textit{k}-mer similarity & with \texttt{--threads} parameter & https://github.com/isovic/graphmap & \cite{sovic2016fast}\\
Minimap & minimizer similarity & with \texttt{-t} parameter & https://github.com/lh3/minimap & \cite{li2016minimap}\\
\hline
\multicolumn{5}{l}{Note: Both GraphMap and Minimap also have read mapping functionality.}
\end{tabularx}
\vspace{5pt}
\caption{State-of-the-art assembly tools.}\label{table:step3}
\begin{tabularx}{\textwidth}{p{1.2cm} p{3.3cm} p{6.0cm} X Z}
\hline
Tool & Strategy & Multi-threading Support & Source & Reference\\ 
\hline
Canu & OLC with error correction & auto configuration & https://github.com/marbl/canu & \cite{koren2017canu}\\
Miniasm & OLC without error correction & no support & https://github.com/lh3/miniasm & \cite{li2016minimap}\\
\hline
\end{tabularx}
\end{table*}
\vspace{-5pt}
\subsection{Read-to-Read Overlap Finding}\label{sec:overlapfinding}
Previous genome assembly methods designed for accurate and short reads (\ie de Bruijn graph (DBG) approach \cite{pevzner2001eulerian,compeau2011apply}) are not suitable for nanopore reads because of the high error rates of the current nanopore sequencing devices \cite{koren2013reducing,de2017sequencer,magi2017nanopore,chu2016innovations}. Instead, overlap-layout-consensus (OLC) algorithms \cite{li2012comparison} are used for nanopore sequencing reads since they perform better with longer, error-prone reads. OLC-based assembly algorithms start with finding the read-to-read overlaps, which is the second step of the pipeline. Read-to-read overlap is defined to be a common sequence between two reads \cite{chu2016innovations}. GraphMap \cite{sovic2016fast} and Minimap \cite{li2016minimap} are the commonly-used state-of-the-art tools for this step (Table~\ref{table:step2}).
\vspace{-5pt}
\subsubsection*{GraphMap}
GraphMap first partitions the entire read dataset into \textit{k}-length substrings (i.e. \textit{k}-mers), and then creates a hash table. GraphMap uses gapped \textit{k}-mers, \ie k-mers that can contain insertions or deletions (indels) \cite{sovic2016fast, burkhardt2003better}. In the hash table, for each \textit{k}-mer entry, three pieces of information are stored: 1) {k}-mer string, 2) the index of the read, and 3) the position in the read where the corresponding \textit{k}-mer comes from. GraphMap detects the overlaps by finding the \textit{k}-mer similarity between any two given reads. Due to this design, GraphMap is a highly sensitive and accurate tool for error-prone long reads. It is a command-line tool written in C++. GraphMap is used for both 1) read-to-read overlap finding with the \texttt{graphmap owler} command and 2) read mapping with the \texttt{graphmap align} command. 
\vspace{-5pt}
\subsubsection*{Minimap}
Minimap also partitions the entire read dataset into \textit{k}-mers, but instead of creating a hash table for the full set of \textit{k}-mers, it finds the minimum representative set of \textit{k}-mers, called \emph{minimizers}, and creates a hash table with only these minimizers. Minimap finds the overlaps between two reads by finding minimizer similarity. The goals of using minimizers are to 1) reduce the storage requirement of the tool by storing fewer \textit{k}-mers and 2) accelerate the overlap finding process by reducing the search span. Minimap also sorts \textit{k}-mers for cache efficiency. Minimap is fast and cache-efficient, and it does not lose any sensitivity by storing minimizers since the chosen minimizers can represent the whole set of \textit{k}-mers. Minimap is a command-line tool written in C. Like GraphMap, it can both 1) find overlaps between two read sets and 2) map a set of reads to a full genome.
\vspace{-15pt}
\subsection{Genome Assembly}\label{sec:assembly}
After finding the read-to-read overlaps, OLC-based assembly algorithms generate an \emph{overlap graph}. Genome assembly is performed by traversing this graph, producing the layout of the reads and then constructing a draft assembly. Canu \cite{koren2017canu} and Miniasm \cite{li2016minimap} are the commonly-used error-prone long-read assemblers (Table~\ref{table:step3}).
\vspace{-10pt}
\subsubsection*{Canu}
Canu performs error-correction as the initial step of its own pipeline. It finds the overlaps of the raw uncorrected reads and uses them for the error-correction. The purpose of error-correction is to improve the accuracy of the bases in the reads \cite{koren2017canu,canu-webpage}. After the error-correction step, Canu finds overlaps between corrected reads and constructs a draft assembly after an additional trimming step. However, error-correction is a computationally expensive step. In its own pipeline, Canu implements its own read-to-read overlap finding tool such that the users do \emph{not} need to perform that step explicitly before running Canu. Most of the steps in the Canu pipeline are multi-threaded. Canu detects the resources that are available in the computer before starting its pipeline and automatically assigns number of threads, number of processes and amount of memory based on the available resources and the assembled genome's estimated size.
\vspace{-10pt}
\subsubsection*{Miniasm}
Miniasm skips the error-correction step since it is computationally expensive. It constructs a draft assembly from the uncorrected read overlaps computed in the previous step. Although Miniasm lowers computational cost and thus accelerates and simplifies assembly by doing so, the accuracy of the draft assembly depends directly on the accuracy of the uncorrected basecalled reads. Thus, further polishing may be necessary for these draft assemblies. Miniasm does not support multi-threading.
\vspace{-10pt}
\begin{table*}[b]
\vspace{5pt}
\caption{State-of-the-art read mapping tools.}\label{table:step4}
\begin{tabularx}{\textwidth}{p{1.5cm} p{4.5cm} p{4.5cm} l Z}
\hline
Tool & Strategy & Multi-threading Support & Source & Reference\\ 
\hline
BWA-MEM & Burrows-Wheeler Transform (BWT) & with \texttt{-t} parameter & http://bio-bwa.sourceforge.net & \cite{li2013aligning}\\
GraphMap & \textit{k}-mer similarity & with \texttt{--threads} parameter & https://github.com/isovic/graphmap & \cite{sovic2016fast}\\
Minimap & minimizer similarity & with \texttt{-t} parameter & https://github.com/lh3/minimap & \cite{li2016minimap}\\
\hline
\end{tabularx}
\vspace{5pt}
\caption{State-of-the-art polishing tools.}\label{table:step5}
\begin{tabularx}{\textwidth}{p{1.5cm} p{4.5cm} p{4.5cm} l Z}
\hline
Tool & Strategy & Multi-threading Support & Source & Reference\\ 
\hline
Nanopolish & Hidden Markov Model & with \texttt{--threads} and \texttt{-P} parameters & https://github.com/jts/nanopolish & \cite{loman2015complete}\\
Racon & Partial order alignment graph & with \texttt{--threads} parameter & https://github.com/isovic/racon & \cite{vaser2017fast}\\
\hline
\end{tabularx}
\vspace{-5pt}
\end{table*}
\subsection{Read Mapping and Polishing}\label{sec:polish}
In order to increase the accuracy of the assembly, especially for the rapid assembly methods like Miniasm, which do not have the error-correction step, further polishing may be required. Polishing, \ie post-assembly error-correction, improves the accuracy of the draft assembly by mapping the reads to the assembly and changing the assembly to increase local similarity with the reads \cite{loman2015complete,vaser2017fast,de2017sequencer}. The first step of polishing is mapping the basecalled reads to the generated draft assembly from the previous step. One of the most commonly-used long read mappers for nanopore data is BWA-MEM \cite{li2013aligning}. Read-to-read overlap finding tools, GraphMap and Minimap (Section~\ref{sec:overlapfinding}), can also be used for this step, since they also have a read mapping mode (Table~\ref{table:step4}).\par
After aligning the basecalled reads to the draft assembly, the final polishing of the assembly can be performed with Nanopolish \cite{loman2015complete} or Racon \cite{vaser2017fast} (Table~\ref{table:step5}).
\vspace{-10pt}
\subsubsection*{Nanopolish}
Nanopolish uses the raw signal data of reads along with the mappings from the previous step to improve the assembly base quality by evaluating and maximizing the probabilities for each base with a Hidden Markov Model-based approach \cite{loman2015complete}. It can increase the accuracy of the draft assembly by correcting the homopolymer-rich parts of the genome. Although this approach can increase the accuracy significantly, it is computationally expensive, and thus time consuming. Nanopolish developers recommend BWA-MEM as the read mapper before running Nanopolish \cite{nanopolishgithub}.
\vspace{-10pt}
\subsubsection*{Racon}
Racon constructs partial order alignment graphs \cite{lee2002multiple,vaser2017fast} in order to find a consensus sequence between the reads and the draft assembly. After dividing the sequence into segments, Racon tries to find the best alignment to increase the accuracy of the draft assembly. Racon is a fast polishing tool, but it does not promise a high increase in accuracy as Nanopolish promises. However, multiple iterations of Racon runs or a combination of Racon and Nanopolish runs can improve accuracy significantly. Racon developers recommend Minimap as the read mapper to use before running Racon, since Minimap is both fast and sensitive \cite{vaser2017fast}.

\vspace{-15pt}
\section{Experimental Methodology} \label{sec:methodology}
\vspace{-2pt}
\subsection{Dataset}
In this work, we use \textit{Escherichia coli} genome data as the test case, sequenced using the MinION with an R9 flowcell \cite{datawebpage}.\par
MinION sequencing has two types of workflows. In the 1D workflow, only the template strand of the double-stranded DNA is sequenced. In contrast, in the 2D workflow, with the help of a hairpin ligation, both the template and complement strands pass through the pore and are sequenced. After the release of R9 chemistry, 1D data became very usable in contrast to previous chemistries. Thus, we perform the analysis of the tools on 1D data.\par
MinION outputs one file in the \emph{fast5} format for each read. The fast5 file format is a hierarchical data format, capable of storing both raw signal data and basecalled data returned by Metrichor. This dataset includes 164,472 reads, \ie fast5 files. Since all these files include both raw signal data and basecalled reads, we can use this dataset for both 1) using the local basecallers to convert raw signal data into the basecalled reads and 2) using the already basecalled reads by Metrichor.
\vspace{-11pt}
\subsection{Evaluation Systems}
In this work, for accuracy and performance evaluations of different tools, we use three separate systems with different specifications. We use the first computer in the first part of the analysis, accuracy analysis. We use the second and third computers in the second part of the analysis, performance analysis, to compare the scalability of the analyzed tools in the two machines with different specifications (Table~\ref{table:specification}).\par
\begin{table*}[t]
\caption{Specifications of evaluation systems.}\label{table:specification}
\begin{threeparttable}
\begin{tabularx}{\textwidth}{l p{2.5cm} p{4.5cm} p{3.4cm} Y}
\hline
Name & Model & CPU Specifications & Main Memory Specifications & NUMA\tnote{*} Specifications\\ 
\hline
System 1 & \specialcel{40-core Intel\textsuperscript \textregistered Xeon\textsuperscript \textregistered\\E5-2630 v4 CPU\\@ 2.20GHz} & \specialcel{20 physical cores\\2 threads per core\\40 logical cores with hyper-threading\tnote{**}} &  \specialcel{128GB DDR4\\2 channels, 2 ranks/channel\\Speed: 2400MHz} & 2 NUMA nodes, each with 10 physical cores, 64GB of memory and an 25MB of last level cache (LLC)\\
& & & & \\
\specialcel{System 2\\(\emph{desktop})} & \specialcel{8-core Intel\textsuperscript \textregistered Core\\i7-2600 CPU\\@ 3.40GHz} & \specialcel{4 physical cores\\2 threads per core\\8 logical cores with hyper-threading\tnote{**}} & \specialcel{16GB DDR3\\2 channels, 2 ranks/channel\\Speed: 1333MHz} & 1 NUMA node, with 4 physical cores, 16GB of memory and an 8MB of LLC\\
& & & & \\
\specialcel{System 3\\(\emph{big-mem})} & \specialcel{80-core Intel\textsuperscript \textregistered Xeon\textsuperscript \textregistered\\E7-4850 CPU\\@ 2.00GHz} & \specialcel{40 physical cores\\2 threads per core\\80 logical cores with hyper-threading\tnote{**}} & \specialcel{1TB DDR3\\8 channels, 4 ranks/channel\\Speed: 1066MHz} & 4 NUMA nodes, each with 10 physical cores, 256GB of memory and an 24MB of LLC\\
\hline
\end{tabularx}
\begin{tablenotes}
\item[*] NUMA (Non-Uniform Memory Access) is a computer memory design, where a processor accesses its local memory faster (\ie with lower latency) than a non-local memory (\ie memory local to another processor in another NUMA node). A NUMA node is composed of the local memory and the CPU cores (See Observation 6 in Section~\ref{sec:basecallresults} for detail).
\item[**] Hyper-threading is Intel's simultaneous multithreading (SMT) implementation (See Observation 5 in Section~\ref{sec:basecallresults} for detail).
\end{tablenotes}
\end{threeparttable}
\end{table*}
We choose the first system for evaluation since it has a larger memory capacity than a usual server and, with the help of a large number of cores, the tasks can be parallelized easily in order to get the output data quickly. We choose the second system, called \emph{desktop}, since it represents a commonly-used desktop server. We choose the third system, called \emph{big-mem}, because of its large memory capacity. This \emph{big-mem} system can be useful for those who would like to get results more quickly.
\vspace{-10pt}
\subsection{Accuracy Metrics}
We compare each draft assembly generated after the assembly step and each improved assembly generated after the polishing step with the reference genome, by using the \texttt{dnadiff} command under the MUMmer package \cite{mummergithub}. We use six metrics to measure accuracy, as defined in Table~\ref{table:accuracymetrics}: 1) number of bases in the assembly, 2) number of contigs, 3) average identity, 4) coverage, 5) number of mismatches, and 6) number of indels.
\vspace{-10pt}
\subsection{Performance Metrics}
We analyze the performance of each tool by running the associated command-line of each tool with the \texttt{/usr/bin/time -v} command. We use four metrics to quantify performance as defined in Table~\ref{table:perfmetrics}: 1) wall clock time, 2) CPU time, 3) peak memory usage, and 4) parallel speedup.
\begin{table*}[b]
\vspace{8pt}
\caption{Accuracy metrics.}\label{table:accuracymetrics}
\begin{tabularx}{\textwidth}{p{2.8cm} p{10.0cm} Z}
\hline
Metric Name & Definition & Preferred Values\\ 
\hline
Number of bases & Total number of bases in the assembly & $\simeq$ Length of reference genome\\
Number of contigs & Total number of segments in the assembly & Lower ($\simeq$ 1)\\
Average identity & Percentage similarity between the assembly and the reference genome & Higher ($\simeq$ 100\%)\\
Coverage & Ratio of the number of aligned bases in the reference genome to the length of reference genome & Higher ($\simeq$ 100\%)\\
Number of mismatches & Total number of single-base differences between the assembly and the reference genome & Lower ($\simeq$ 0)\\
Number of indels & Total number of insertions and deletions between the assembly and the reference genome & Lower ($\simeq$ 0)\\
\hline
\end{tabularx}
\vspace{1pt}
\caption{Performance metrics.}\label{table:perfmetrics}
\begin{threeparttable}
\begin{tabularx}{\textwidth}{p{2.8cm} p{10.0cm} Z}
\hline
Metric Name & Definition & Preferred Values\\ 
\hline
Wall clock time & Elapsed time from the start of a program to the end & Lower\\
CPU time & Total amount of time the CPU spends in user mode (\ie to run the program's code) and kernel mode (\ie to execute system calls made by the program)\tnote{*}& Lower\\
Peak memory usage & Maximum amount of memory used by a program during its whole lifetime & Lower\\
Parallel speedup & Ratio of the time to run a program with 1 thread to the time to run it with N threads & Higher\\
\hline
\end{tabularx}
\begin{tablenotes}
\item[*] If wall clock time < CPU time for a specific program, it means that the program runs in parallel. 
\end{tablenotes}
\end{threeparttable}
\vspace{-5pt}
\end{table*}
\begin{table*}[t]
\caption{Versions, commands to execute, and outputs for each analyzed tool.}\label{table:overalltools}
\begin{threeparttable}
\begin{tabularx}{\textwidth}{l p{10cm} Z}
\hline
& Command\tnote{*} & Output\\
\hline
\multicolumn{3}{l}{\textcolor{mypink}{\textbf{Basecalling Tools}}}\\
\hline
Nanonet--v2.0 & \texttt{nanonetcall fast5\_dir/ -{}-jobs N -{}-chemistry r9} & \texttt{reads.fasta}\\
Scrappie--v1.0.1& \specialcel{\texttt{(1)export OMP\_NUM\_THREADS=N}\\\texttt{(2)scrappie events -{}-segmentation Segment\_Linear:split\_hairpin}\\\texttt{\textcolor{white}{(2)}fast5\_dir/ ...}} & \specialcel{\\\\\texttt{reads.fasta}}\\
Nanocall--v0.7.4& \texttt{nanocall -t N fast5\_dir/} & \texttt{reads.fasta}\\
DeepNano--e8a621e & \texttt{python basecall.py --directory fast5\_dir/ -{}-chemistry r9} & \texttt{reads.fasta}\\
\hline
\multicolumn{3}{l}{\textcolor{myblue}{\textbf{Read-to-Read Overlap Finding Tools}}}\\
\hline
GraphMap--v0.5.2 & \texttt{graphmap owler -L paf -t N -r reads.fasta -d reads.fasta} & \texttt{overlaps.paf}\\
Minimap--v0.2 & \texttt{minimap -Sw5 -L100 -m0 -tN reads.fasta reads.fasta} & \texttt{overlaps.paf}\\
\hline
\multicolumn{3}{l}{\textcolor{mygreen}{\textbf{Assembly Finding Tools}}}\\
\hline
Canu--v1.6 & \texttt{canu -p ecoli -d canu-ecoli genomeSize=4.6m -nanopore-raw reads.fasta} & \specialcel{\\\texttt{draftAssembly.fasta}}\\
Miniasm--v0.2 & \texttt{miniasm -f reads.fasta overlaps.paf} & \texttt{draftAssembly.gfa --> draftAssembly.fasta}\\
\hline
\multicolumn{3}{l}{\textcolor{myorange}{\textbf{Read Mapping Tools}}}\\
\hline
BWA-MEM--0.7.15 & \specialcel{\texttt{(1)bwa index draftAssembly.fasta}\\ \texttt{(2)bwa mem -x ont2d -t N draftAssembly.fasta reads.fasta}} & \specialcel{\\\texttt{mappings.sam -->}\\\texttt{\textcolor{white}{ --> }mappings.bam}}\\
Minimap--v0.2 & \texttt{minimap -tN draftAssembly.fasta reads.fasta} & \texttt{mappings.paf}\\
\hline
\multicolumn{3}{l}{\textcolor{mypurple}{\textbf{Polishing Tools}}}\\
\hline
Nanopolish--v0.7.1 & \specialcel{\texttt{(1)python nanopolish\_makerange.py draftAssembly.fasta | parallel -P M}\\ \texttt{(2)nanopolish variants -{}-consensus polished.\verb|{1}|.fa -w \verb|{1}|}\\\texttt{\textcolor{white}{(2)}-r reads.fasta -b mappings.bam -g draftAssembly.fasta -t N}\\\texttt{(3)python nanopolish\_merge.py polished.*.fa}} & \specialcel{\\\\\\\texttt{polished.fasta}}\\
Racon--v0.5.0 & \texttt{racon (--sam) --bq -1 -t N reads.fastq mappings.paf/(mappings.sam) draftAssembly.fasta} & \specialcel{\\\texttt{polished.fasta}}\\
\hline
\end{tabularx}
\begin{tablenotes}
\item[*] N corresponds to the number of threads and M corresponds to the number of parallel jobs.
\end{tablenotes}
\end{threeparttable}
\end{table*}

\vspace{-15pt}
\enlargethispage{\baselineskip}
\section{Results and Analysis} \label{sec:results-analysis}
\begin{table*}[t]
\caption{Accuracy analysis results using different tools for the first three steps of the pipeline.}\label{table:accuracyfirst3}
\begin{tabularx}{\textwidth}{l | *{6}{Z}}
\hline
& \specialcell{Number of\\Bases} & \specialcell{Number of\\Contigs} & \specialcell{Identity\\(\%)} & \specialcell{Coverage\\(\%)} & \specialcell{Number of\\Mismatches} & \specialcell{Number of\\Indels}\\ 
\hline
\begin{tabular}{p{0.3cm} | p{1.1cm} l p{1.1cm} r p{1.4cm}}1 & \textcolor{mypink}{Metrichor} & + & --- & + & \textcolor{mygreen}{Canu}\end{tabular} & 4,609,499 & 1 & 98.05 & 99.92 & 12,378 & 76,990\\
\begin{tabular}{p{0.3cm} | p{1.1cm} l p{1.1cm} r p{1.1cm}}2 & \textcolor{mypink}{Metrichor} & + & \textcolor{myblue}{Minimap} & + & \textcolor{mygreen}{Miniasm}\end{tabular} & 4,402,675 & 1 & 87.71 & 94.85 & 249,096 & 372,704\\
\begin{tabular}{p{0.3cm} | p{1.1cm} l p{1.1cm} r p{1.1cm}}3 & \textcolor{mypink}{Metrichor} & + & \textcolor{myblue}{GraphMap} & + & \textcolor{mygreen}{Miniasm}\end{tabular}  & 4,500,155 & 2 & 86.22 & 96.95 & 237,747 & 360,199\\
\hline
\begin{tabular}{p{0.3cm} | p{1.1cm} l p{1.1cm} r p{1.1cm}}4 & \textcolor{mypink}{Nanonet} & + & --- & + & \textcolor{mygreen}{Canu}\end{tabular} & 4,581,728 & 1 & 97.92 & 99.97 & 11,971 & 83,248\\
\begin{tabular}{p{0.3cm} | p{1.1cm} l p{1.1cm} r p{1.1cm}}5 & \textcolor{mypink}{Nanonet} & + & \textcolor{myblue}{Minimap} & + & \textcolor{mygreen}{Miniasm}\end{tabular} & 4,350,175 & 1 & 85.50 & 92.76 & 237,518 & 394,852\\
\begin{tabular}{p{0.3cm} | p{1.1cm} l p{1.1cm} r p{1.1cm}}6 & \textcolor{mypink}{Nanonet} & + & \textcolor{myblue}{GraphMap} & + & \textcolor{mygreen}{Miniasm}\end{tabular} & 4,272,545 & 1 & 85.36 & 91.16 & 232,748 & 389,968\\
\hline
\begin{tabular}{p{0.3cm} | p{1.1cm} l p{1.1cm} r p{1.1cm}}7 & \textcolor{mypink}{Scrappie} & + & --- & + & \textcolor{mygreen}{Canu}\end{tabular}& 4,614,149 & 1 & 98.46 & 99.90 & 6,777 & 63,597\\
\begin{tabular}{p{0.3cm} | p{1.1cm} l p{1.1cm} r p{1.1cm}}8 & \textcolor{mypink}{Scrappie} & + & \textcolor{myblue}{Minimap} & + & \textcolor{mygreen}{Miniasm}\end{tabular} & 4,877,399 & 8 & 86.94 & 90.04 & 184,669 & 363,025\\
\begin{tabular}{p{0.3cm} | p{1.1cm} l p{1.1cm} r p{1.1cm}}9 & \textcolor{mypink}{Scrappie} & + & \textcolor{myblue}{GraphMap} & + & \textcolor{mygreen}{Miniasm}\end{tabular} & 4,368,417 & 1 & 86.78 & 89.86 & 189,192 & 372,245\\
\hline
\begin{tabular}{p{0.3cm} | p{1.1cm} l p{1.1cm} r p{1.1cm}}10& \textcolor{mypink}{Nanocall} & + & --- & + & \textcolor{mygreen}{Canu}\end{tabular} & 1,299,808 & 86 & 93.33 & 28.93 & 21,985 & 61,217\\
\begin{tabular}{p{0.3cm} | p{1.1cm} l p{1.1cm} r p{1.1cm}}11 & \textcolor{mypink}{Nanocall} & + & \textcolor{myblue}{Minimap} & + & \textcolor{mygreen}{Miniasm}\end{tabular} & 4,492,964 & 5 & 80.52 & 42.92 & 177,589 & 221,092\\
\begin{tabular}{p{0.3cm} | p{1.1cm} l p{1.1cm} r p{1.1cm}}12 & \textcolor{mypink}{Nanocall} & + & \textcolor{myblue}{GraphMap} & + & \textcolor{mygreen}{Miniasm}\end{tabular} & 4,429,390 & 3 & 80.51 & 41.32 & 171,455 & 213,435\\
\hline
\begin{tabular}{p{0.3cm} | p{1.1cm} l p{1.1cm} r p{1.1cm}}13 & \textcolor{mypink}{DeepNano} & + & --- & + & \textcolor{mygreen}{Canu}\end{tabular} & 7,151,596 & 106 & 92.75 & 99.16 & 38,803 & 211,551\\
\begin{tabular}{p{0.3cm} | p{1.1cm} l p{1.1cm} r p{1.1cm}}14 & \textcolor{mypink}{DeepNano} & + & \textcolor{myblue}{Minimap} & + & \textcolor{mygreen}{Miniasm}\end{tabular} & 4,252,525 & 1 & 82.38 & 65.00 & 199,122 & 335,761\\
\begin{tabular}{p{0.3cm} | p{1.1cm} l p{1.1cm} r p{1.1cm}}15 & \textcolor{mypink}{DeepNano} & + & \textcolor{myblue}{GraphMap} & + & \textcolor{mygreen}{Miniasm}\end{tabular}  & 4,251,548 & 1 & 82.39 & 64.92 & 197,914 & 335,064\\
\hline
\end{tabularx}
\vspace{3pt}
\caption{Performance analysis results for the first three steps of the pipeline.}\label{table:perffirst3}
\begin{threeparttable}
\begin{tabular}{l |r r r| r r r | r r r}
\hline
& \multicolumn{3}{c|}{\textcolor{mypink}{\specialcellll{\textbf{Step 1:}\\Basecaller}}} &  \multicolumn{3}{c|}{\textcolor{myblue}{\specialcellll{\textbf{Step 2:}\\Read-to-Read Overlap Finder}}} & \multicolumn{3}{c}{\textcolor{mygreen}{\specialcellll{\textbf{Step 3:}\\Assembly}}}\\
\cline{2-10}
& \specialcell{Wall\\Clock\\Time\\(h:m:s)} & \specialcell{CPU Time\\(h:m:s)} & \specialcell{Memory\\Usage\\(GB)} & \specialcell{Wall\\Clock\\Time\\(h:m:s)} & \specialcell{CPU Time\\(h:m:s)} & \specialcell{Memory\\Usage\\(GB)} & \specialcell{Wall\\Clock\\Time\\(h:m:s)} & \specialcell{CPU Time\\(h:m:s)} & \specialcell{Memory\\Usage\\(GB)}\\ 
\hline
\begin{tabular}{p{0.3cm} | p{1.1cm} l p{1.1cm} r p{1.4cm}}1 & \textcolor{mypink}{Metrichor} & + & --- & + & \textcolor{mygreen}{Canu}\end{tabular} & \multirow{3}{*}{---\tnote{*}} & \multirow{3}{*}{---\tnote{*}} & \multirow{3}{*}{---\tnote{*}} & --- & --- & --- & 44:12:31 & 502:18:56 & 5.76\\
\begin{tabular}{p{0.3cm} | p{1.1cm} l p{1.1cm} r p{1.1cm}}2 & \textcolor{mypink}{Metrichor} & + & \textcolor{myblue}{Minimap} & + & \textcolor{mygreen}{Miniasm}\end{tabular} & & & & 2:15 & 41:37 & 12.30 & 1:09 & 1:09 & 1.96\\
\begin{tabular}{p{0.3cm} | p{1.1cm} l p{1.1cm} r p{1.1cm}}3 & \textcolor{mypink}{Metrichor} & + & \textcolor{myblue}{GraphMap} & + & \textcolor{mygreen}{Miniasm}\end{tabular} & & & & 6:14 & 1:52:57 & 56.58 & 1:05 & 1:05 & 1.82\\
\hline
\begin{tabular}{p{0.3cm} | p{1.1cm} l p{1.1cm} r p{1.1cm}}4 & \textcolor{mypink}{Nanonet} & + & --- & + & \textcolor{mygreen}{Canu}\end{tabular} & \multirow{3}{*}{17:52:42} & \multirow{3}{*}{714:21:45} & \multirow{3}{*}{1.89} & --- & --- & --- & 11:32:40 & 129:07:16 & 5.27\\
\begin{tabular}{p{0.3cm} | p{1.1cm} l p{1.1cm} r p{1.1cm}}5 & \textcolor{mypink}{Nanonet} & + & \textcolor{myblue}{Minimap} & + & \textcolor{mygreen}{Miniasm}\end{tabular} & & & & 1:13 & 18:55 & 9.45 & 33 & 33 & 0.69\\
\begin{tabular}{p{0.3cm} | p{1.1cm} l p{1.1cm} r p{1.1cm}}6 & \textcolor{mypink}{Nanonet} & + & \textcolor{myblue}{GraphMap} & + & \textcolor{mygreen}{Miniasm}\end{tabular} & & & & 3:18 & 48:27 & 29.16 & 32 & 32 & 0.65\\
\hline
\begin{tabular}{p{0.3cm} | p{1.1cm} l p{1.1cm} r p{1.1cm}}7 & \textcolor{mypink}{Scrappie} & + &--- & + & \textcolor{mygreen}{Canu}\end{tabular} & \multirow{3}{*}{3:11:41} & \multirow{3}{*}{126:19:06} & \multirow{3}{*}{13.36} & --- & --- & --- & 33:47:41 & 385:51:23 & 5.75\\
\begin{tabular}{p{0.3cm} | p{1.1cm} l p{1.1cm} r p{1.1cm}}8 & \textcolor{mypink}{Scrappie} & + & \textcolor{myblue}{Minimap} & + & \textcolor{mygreen}{Miniasm}\end{tabular} & & & & 2:52 & 1:10:26 & 12.40 & 1:29 & 1:29 & 1.98\\
\begin{tabular}{p{0.3cm} | p{1.1cm} l p{1.1cm} r p{1.1cm}}9 & \textcolor{mypink}{Scrappie} & + & \textcolor{myblue}{GraphMap} & + & \textcolor{mygreen}{Miniasm}\end{tabular} & & & & 7:26 & 2:16:02 & 38.31 & 1:23 & 1:23 & 1.87\\
\hline
\begin{tabular}{p{0.3cm} | p{1.1cm} l p{1.1cm} r p{1.1cm}}10 & \textcolor{mypink}{Nanocall} & + & --- & + & \textcolor{mygreen}{Canu}\end{tabular} & \multirow{3}{*}{47:04:53} & \multirow{3}{*}{1857:37:56} & \multirow{3}{*}{37.73} & --- & --- & --- & 1:35:23 & 27:58:29 & 3.77\\
\begin{tabular}{p{0.3cm} | p{1.1cm} l p{1.1cm} r p{1.1cm}}11 & \textcolor{mypink}{Nanocall} & + & \textcolor{myblue}{Minimap} & + & \textcolor{mygreen}{Miniasm}\end{tabular} & & & & 1:15 & 16:08 & 12.19 & 20 & 20 & 0.47\\
\begin{tabular}{p{0.3cm} | p{1.1cm} l p{1.1cm} r p{1.1cm}}12 & \textcolor{mypink}{Nanocall} & + & \textcolor{myblue}{GraphMap} & + & \textcolor{mygreen}{Miniasm}\end{tabular} & & & & 5:14 & 1:09:04 & 56.78 & 16 & 16 & 0.30\\
\hline
\begin{tabular}{p{0.3cm} | p{1.1cm} l p{1.1cm} r p{1.1cm}}13 & \textcolor{mypink}{DeepNano} & + & --- & + & \textcolor{mygreen}{Canu}\end{tabular} & \multirow{3}{*}{23:54:34} & \multirow{3}{*}{811:14:29} & \multirow{3}{*}{8.38} & --- & --- & --- & 1:15:48 & 17:31:07 & 3.61\\
\begin{tabular}{p{0.3cm} | p{1.1cm} l p{1.1cm} r p{1.1cm}}14 & \textcolor{mypink}{DeepNano} & + & \textcolor{myblue}{Minimap} & + & \textcolor{mygreen}{Miniasm}\end{tabular} & & & & 1:50 & 24:30 & 11.71 & 1:03 & 1:03 & 1.31\\
\begin{tabular}{p{0.3cm} | p{1.1cm} l p{1.1cm} r p{1.1cm}}15 & \textcolor{mypink}{DeepNano} & + & \textcolor{myblue}{GraphMap} & + & \textcolor{mygreen}{Miniasm}\end{tabular} & & & & 5:18 & 1:17:06 & 54.64 & 58 & 58 & 1.10\\
\hline
\end{tabular}
\begin{tablenotes}
\item[*] We cannot get the performance metrics for Metrichor since its source code is not available for us to run the tool by ourselves.
\end{tablenotes}
\end{threeparttable}
\vspace{-10pt}
\end{table*}
In this section, we present our results obtained by analyzing the performance of different tools for each step in the genome assembly pipeline using nanopore sequence data in terms of accuracy and performance, using all the metrics we provide in Table~\ref{table:accuracymetrics} and Table~\ref{table:perfmetrics}. Additionally, Table~\ref{table:overalltools} shows the tool version, the executed command, and the output of each analyzed tool. We divide our analysis into three main parts.\par
In the first part of our analysis, we examine the first three steps of the pipeline (\emph{cf.} Figure~\ref{fig:pipeline}). To this end, we first execute each basecalling tool (\ie one of Nanonet, Scrappie, Nanocall or DeepNano). Since Metrichor is a cloud-based tool and its source code is not available, we cannot execute Metrichor and get the performance metrics for it. After recording the performance metrics for each basecaller run, we execute either GraphMap or Minimap followed by Miniasm, or Canu itself, and record the performance metrics for each run. We obtain a draft assembly for each combination of these basecalling, read-to-read overlap finding and assembly tools. For each draft assembly, we assess its accuracy by comparing the resulting draft assembly with the existing reference genome. We show the accuracy results in Table~\ref{table:accuracyfirst3}. We show the performance results in Table~\ref{table:perffirst3}. We will refer to these tables in Sections~\ref{sec:basecallresults} --~\ref{sec:assemblyresults}.\par
In the second part of our analysis, we examine the last two steps of the pipeline (\emph{cf.} Figure~\ref{fig:pipeline}). To this end, for each obtained draft assembly, we execute each possible combination of different read mappers (\ie BWA-MEM or Minimap) and different polishers (\ie Nanopolish or Racon), and record the performance metrics for each step (\ie read mapping and polishing). We obtain a polished assembly after each run, and assess its accuracy by comparing it with the existing reference genome. For these two analyses, we use the first system, which has 40 logical cores, and execute each tool using 40 threads, which is the possible maximum number of threads for that particular machine. We show the accuracy results in Table~\ref{table:accuracylast2}. We show the performance results in Table~\ref{table:perflast2}. We will refer to these tables in Section~\ref{sec:polishresults}.\par
In the third part of our analysis, we assess the scalability of all of the tools that have multi-threading support. For this purpose, we use the second and third systems to compare the scalability of these tools on two different system configurations. For each tool, we change the number of threads and observe the corresponding change in speed, memory usage, and parallel speedup. These results are depicted in Figures~\ref{fig:basecallplot} --~\ref{fig:nanopolishplot}, and we will refer to them throughout Sections~\ref{sec:basecallresults} --~\ref{sec:polishresults}.\par
Sections~\ref{sec:basecallresults} --~\ref{sec:polishresults} describe the major observations we make for each of the five steps of the pipeline (\emph{cf.} Figure~\ref{fig:pipeline}) based on our extensive evaluation results.
\vspace{-11pt}
\subsection{Basecalling Tools}\label{sec:basecallresults}
As we discuss in Section~\ref{sec:basecalling}, ONT's basecallers Metrichor, Nanonet and Scrappie, and another basecaller developed by Boza \textit{et al}. (2017), DeepNano, use Recurrent Neural Networks (RNNs) for basecalling whereas Nanocall developed by David \etal (2016) uses Hidden Markov Models (HMM) for basecalling.
\vspace{-10pt}
\subsubsection*{Accuracy}
Using RNNs is a more powerful basecalling approach than using HMMs since an RNN 1) does not make any assumptions about sequence length \cite{sutskever2014sequence} and 2) is not affected by the repeats in the sequence \cite{sutskever2014sequence,david2016nanocall,bovza2017deepnano}. However, it is still challenging to determine the correct length of the homopolymers even with an RNN.\par
In order to compare the accuracy of the analyzed basecallers, we group the accuracy results by each basecalling tool and compare them according to the defined accuracy metrics.\par
According to this analysis and the accuracy results shown in Table~\ref{table:accuracyfirst3}, we make the following key observation.\par
\textbf{Observation 1: }\textit{The pipelines that start with Metrichor, Nanonet, or Scrappie as the basecaller have similar identity and coverage trends among all of the evaluated scenarios (\ie tool combinations for the first three steps), but Scrappie has a lower number of mismatches and indels. However, Nanocall and DeepNano cannot reach these three basecallers' accuracies: they have lower identity and lower coverage.}\par
Since Nanonet is the local version of Metrichor, Nanonet and Metrichor's similar accuracy trends are expected. In addition to the power of the RNN-based approach, Scrappie tries to solve the homopolymer basecalling problem. Although Scrappie is in an early stage of development, it leads to a smaller number of indels than Metrichor or Nanonet. Nanocall's poor accuracy results are due to the simple HMM-based approach it uses. Although DeepNano performs better than Nanocall with the help of its RNN-based approach, it results in a higher number of indels and a lower coverage of the reference genome.
\subsubsection*{Performance}
RNN and HMM are computationally-intensive algorithms. In HMM-based basecalling, the Viterbi algorithm \cite{forney1973viterbi} is used for decoding. The Viterbi algorithm is a sequential technique and its computation cannot currently be parallelized with multithreading. However, in RNN-based basecalling, multiple threads can work on different sections of the neural network and thus RNN computation can be parallelized with multithreading.\par
In order to measure and compare the performance of the selected basecallers, we first compare the recorded wall clock time, CPU time and memory usage metrics of each scenario for the first step of the pipeline. Based on the results provided in Table~\ref{table:perffirst3}, we make the following key observation.\par
\textbf{Observation 2: }\textit{RNN-based Nanonet and DeepNano are 2.6x and 2.3x faster than HMM-based Nanocall, respectively. Although Scrappie is also an RNN-based tool, it is 5.7x faster than Nanonet due to its C implementation as opposed to Nanonet's Python implementation.}\par
\begin{figure*}[t]
\vspace{-4pt}
\centering
\includegraphics[width=15cm,height=11.5cm]{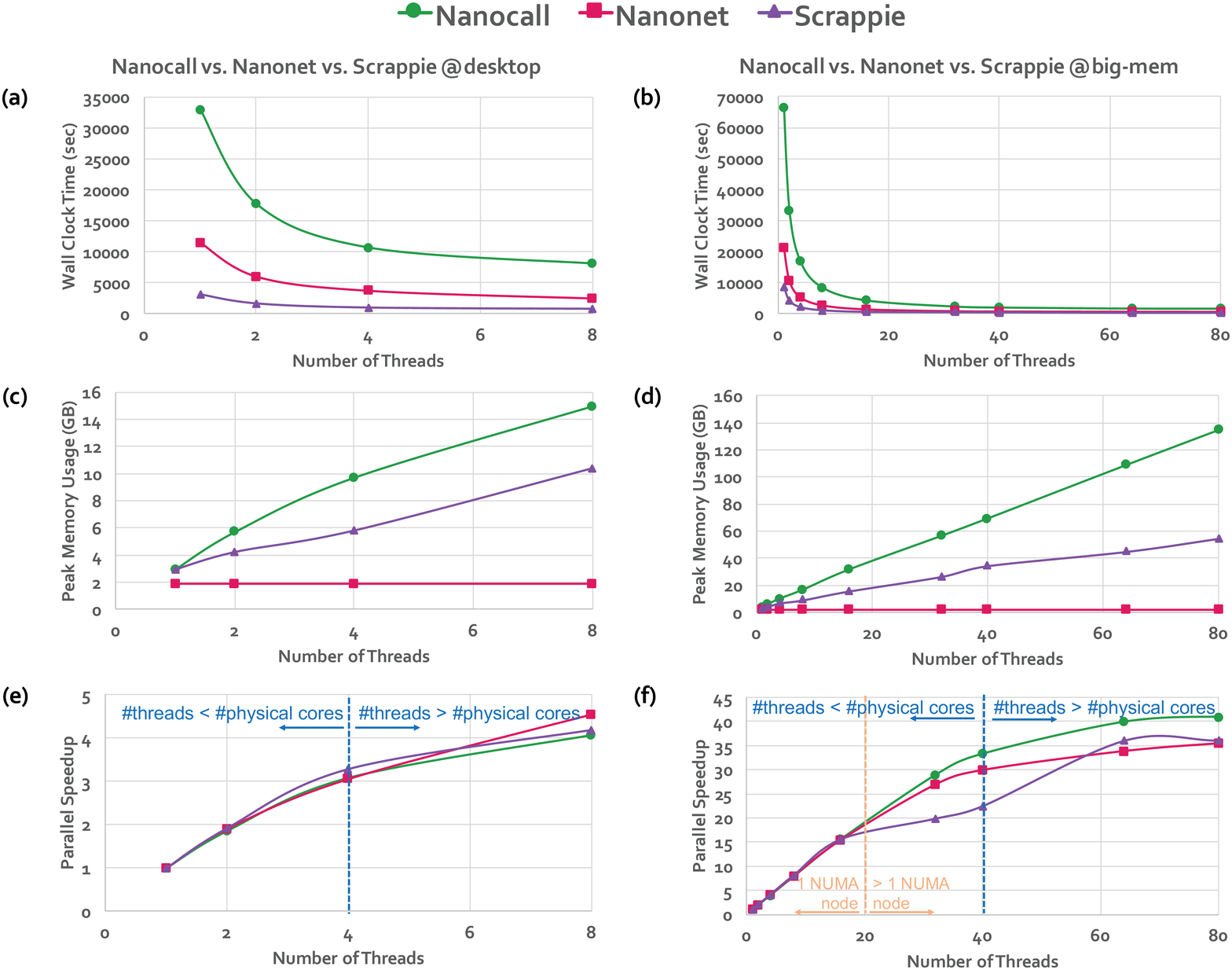}
\caption{Scalability results of Nanocall, Nanonet and Scrappie.}\label{fig:basecallplot}
\raggedright
Wall clock time (a, b), peak memory usage (c, d), and parallel speedup (e, f) results obtained on the \textit{desktop} and \textit{big-mem} systems. The left column (a, c, e) shows the results from the \textit{desktop} system and the right column (b, d, f) shows the results from the \textit{big-mem} system.
\vspace{-5pt}
\end{figure*}

For a deeper understanding of these tools' advantages, disadvantages and bottlenecks, we also perform a scalability analysis for each basecaller by running it on the \textit{desktop} server and the \textit{big-mem} server separately, with 1, 2, 4, 8 (maximum for the \textit{desktop} server), 16, 32, 40, 64 and 80 (maximum for the \textit{big-mem} server) threads, and measuring the performance metrics for each configuration. Metrichor and DeepNano are not included in this analysis because Metrichor is a cloud-based tool and its source code is not available for us to change its number of threads, and DeepNano does not support multi-threading. Figure~\ref{fig:basecallplot} shows the speed, memory usage and parallel speedup results of our evaluations. We make four observations.\par
\textbf{Observation 3: }\textit{When we compare \textit{desktop}'s and \textit{big-mem}'s single thread performance, we observe that \textit{desktop} is approximately 2x faster than \textit{big-mem} (cf. Figure~\ref{fig:basecallplot}a and~\ref{fig:basecallplot}b).}\par
This is mainly because of \textit{desktop}'s higher CPU frequency (see Table 6). It is an indication that all of these three tools are computationally expensive. Larger memory capacity or larger Last-Level Cache (LLC) capacity of \textit{big-mem} cannot make up for the higher CPU speed in \textit{desktop} when there is only one thread.\par
\textbf{Observation 4: }\textit{Scrappie and Nanocall have a linear increase in memory usage when number of threads increases. In contrast, Nanonet has a constant memory usage for all evaluated thread units (cf. Figure~\ref{fig:basecallplot}c and~\ref{fig:basecallplot}d)}.\par
In Scrappie and Nanocall, each thread performs the basecalling for different groups of raw reads. Thus, each thread allocates its own memory space for the corresponding data. This causes the linear increase in memory usage when the level of parallelism increases. In Nanonet, all of the threads share the computation of each read, and thus memory usage is not affected by the amount of thread parallelism.\par
\textbf{Observation 5: }\textit{When the number of threads exceeds the number of physical cores, the simultaneous multithreading overhead prevents continued linear speedup of Nanonet, Scrappie and Nanocall (cf. Figure~\ref{fig:basecallplot}e and~\ref{fig:basecallplot}f).}\par
Simultaneous multithreading (SMT) (\ie running more than one thread per physical core \cite{marr2002hyper,magro2002hyper,tuck2003initial,tullsen1995simultaneous,eggers1997simultaneous,tullsen1996exploiting,yamamoto1995increasing,hirata1992elementary}), or more specifically Intel's hyper-threading (\ie since we use Intel's hyper-threading enabled machines (see Table~\ref{table:specification})) helps to decrease the total runtime but it does \emph{not} provide a linear speedup with the number of threads because of the CPU-intensive workload of Scrappie, Nanocall and Nanonet. If the threads executed are CPU-bound and do not wait for the memory or I/O requests, hyper-threading does not provide linear speedup due to the contention it causes in the shared resources for the computation. This phenomenon has been analyzed extensively in other application domains \cite{marr2002hyper,magro2002hyper,tuck2003initial}.\par
\textbf{Observation 6: }\textit{Data sharing between threads degrades the parallel speedup of Nanonet when cores from multiple NUMA nodes take role in the computation (cf. Figure~\ref{fig:basecallplot}f).}\par
In Nanonet, data is shared between threads and each thread performs different computations on the same data. There are 4 NUMA nodes in \textit{big-mem} (\emph{cf.} Table~\ref{table:specification}), and when data is shared between multiple NUMA nodes, this negatively affects the speedup of Nanonet because accessing the data located in another node (\ie non-local memory) requires longer latency than accessing the data located in local memory. When multiple NUMA nodes start taking role in the computation, Nanocall performs better in terms of scalability since it does \emph{not} require data sharing between different threads.\par
\textbf{Summary.} Based on the observations we make about the analyzed basecalling tools, we conclude that the choice of the tool for this step plays an important role to overcome the high error rates of nanopore sequencing technology. Basecalling with Recurrent Neural Networks (\eg Metrichor, Nanonet, Scrappie) provides higher accuracy and higher speed than basecalling with Hidden Markov Models, and the newest basecaller of ONT, Scrappie, also has the potential to overcome the homopolymer basecalling problem. 
\vspace{-13pt}
\subsection{Read-to-Read Overlap Finding Tools}\label{sec:overlapresults}
As we discuss in Section~\ref{sec:overlapfinding}, GraphMap and Minimap are the commonly-used tools for this step. GraphMap finds the overlaps using \textit{k}-mer similarity, whereas Minimap finds them by using minimizers instead of the full set of \textit{k}-mers. 
\vspace{-15pt}
\subsubsection*{Accuracy}
As done in GraphMap, finding the overlaps with the help of full set of \textit{k}-mers is a highly-sensitive and accurate approach. However, it is also resource-intensive. For this reason, instead of the full set of \textit{k}-mers, Minimap uses a minimum representative set of \textit{k}-mers, minimizers, as an alternative approach for finding the overlaps.\par
In order to compare the accuracy of these two approaches, we categorize the results in Table~\ref{table:accuracyfirst3} based on read-to-read overlap finding tools. In other words, we look at the rows with the same basecaller (\ie red-labeled tools) and same assembler (\ie green-labeled tools) but different read-to-read overlap finder (\ie blue-labeled tools). After that, we compare them according to the defined accuracy metrics. We make the following major observation.\par
\textbf{Observation 7: }\textit{Pipelines with GraphMap or Minimap end up with similar values for identity, coverage, number of indels and mismatches. Thus, either of these read-to-read overlap finding tools can be used in the second step of the nanopore sequencing assembly pipeline to achieve similar accuracy.}\par
Minimap and GraphMap do not have a significantly different effect on the accuracy of the generated draft assemblies. This is because Minimap does not lose any sensitivity by storing minimizers instead of the full set of \textit{k}-mers. 
\vspace{-10pt}
\subsubsection*{Performance}
In order to compare the performance of GraphMap and Minimap, we categorize the results in Table~\ref{table:perffirst3} based on read-to-read overlap finding tools, in a similar way we describe the results in Table~\ref{table:accuracyfirst3} for the accuracy analysis. We also perform a scalability analysis for each of these tools by running them on the \textit{big-mem} server with 1, 2, 4, 8, 16, 32, 40, 64 and 80 threads, and measuring the performance metrics. Because of the high memory usage of GraphMap, data necessary for the tool does not fit in the memory of the \textit{desktop} server and the GraphMap job exits due to a bad memory allocation exception. Thus, we could not perform the scalability analysis of GraphMap in the \textit{desktop} server.\par
\begin{figure}[H]
\vspace{-22pt}
\centering
\includegraphics[width=\columnwidth,height=12.5cm]{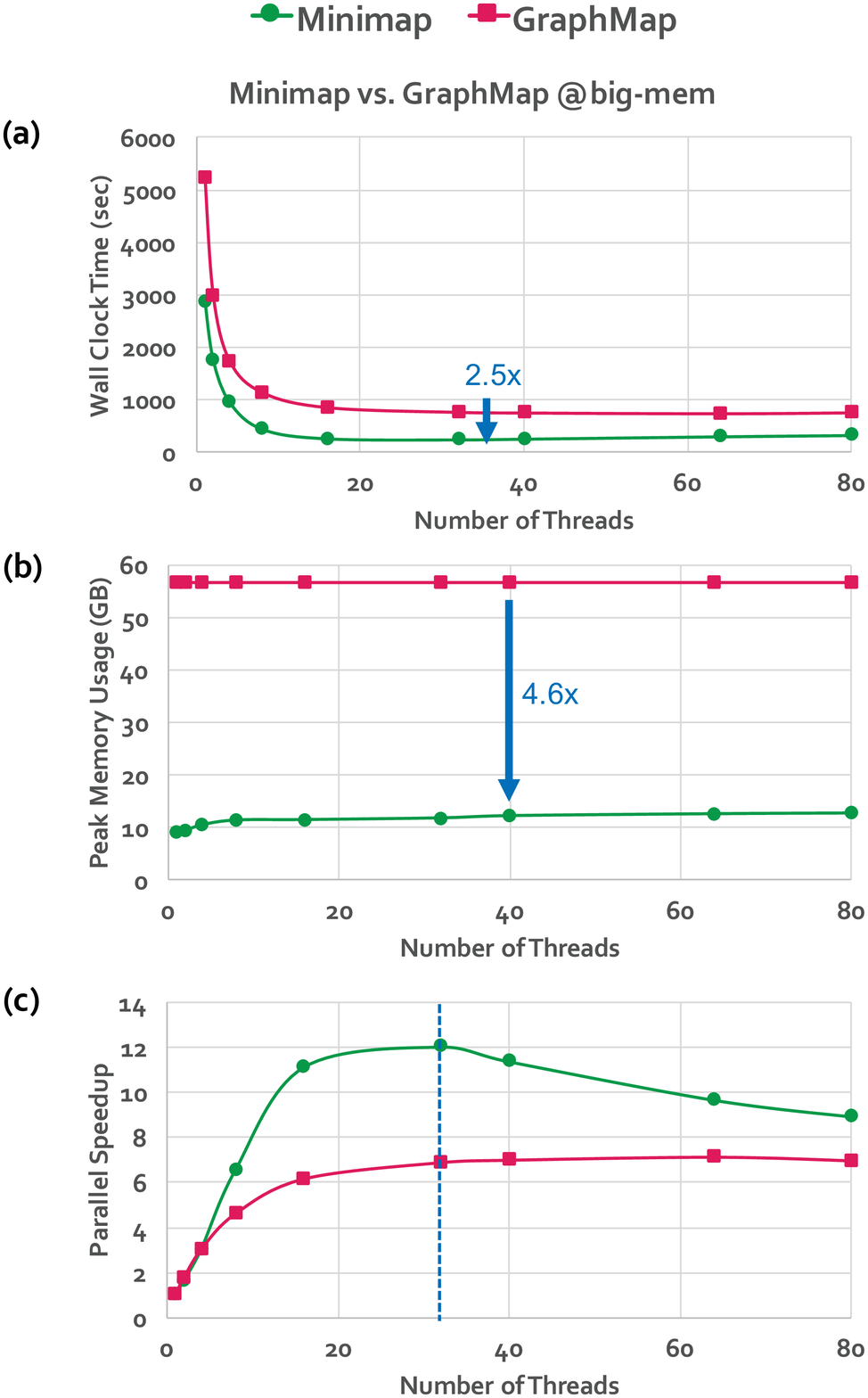}
\caption{Scalability results of Minimap and GraphMap.}\label{fig:overlapplot}
\raggedright
Wall clock time (a), peak memory usage (b), and parallel speedup (c) results obtained on the \textit{big-mem} system.
\vspace{-10pt}
\end{figure}

Figure~\ref{fig:overlapplot} depicts the speed, memory usage and parallel speedup results of the scalability analysis for GraphMap and Minimap. We make the following three observations according to the results from Table~\ref{table:perffirst3} and Figure~\ref{fig:overlapplot}.\par
\textbf{Observation 8: }\textit{The memory usage of both GraphMap and Minimap is dependent on the hash table size but independent of number of threads. Minimap requires 4.6x less memory than GraphMap, on average.}\par
This is mainly because Minimap stores only minimizers instead of all \textit{k}-mers. Storing the full set of \textit{k}-mers in GraphMap requires a larger hash table, and thus higher memory usage than Minimap. The high amount of memory requirement causes GraphMap to not run on our \emph{desktop} system for none of the selected number of thread units.\par
\textbf{Observation 9: }\textit{Minimap is 2.5x faster than GraphMap, on average, across different scenarios in Table~\ref{table:perffirst3}.}\par
Since GraphMap stores all \textit{k}-mers, GraphMap needs to scan its very large dataset while finding the overlaps between two reads. However, in Minimap, the size of dataset that needs to be scanned is greatly shrunk by storing minimizers, as we describe in Observation 8. Thus, Minimap performs much less computation, leading to its 2.5x speedup. Another indication of the different memory usage and its effect on the speed of computation is the Last-Level Cache (LLC) miss rates of these two tools. The LLC miss rate of Minimap is 36\% whereas the LLC miss rate of GraphMap is 55\%. Since the size of data needed by GraphMap is much larger than the LLC size, GraphMap experiences LLC misses more frequently. As a result, GraphMap stalls for longer, waiting for data accesses from main memory, which negatively affects the speed of the tool.\par
\textbf{Observation 10: }\textit{Minimap is more scalable than GraphMap. However, after 32 threads, there is a decrease in the parallel speedup of Minimap (cf. Figure~\ref{fig:overlapplot}c).}\par
Because of its lower computational workload and lower memory usage, we find that Minimap is more scalable than GraphMap. However, in Minimap, threads that finish their work wait for the other active threads to finish their workloads, before starting new work, in order to prevent higher memory usage. Because of this, when the number of threads reaches a high number (\ie 32 in Figure~\ref{fig:overlapplot}c), synchronization overhead greatly increases, causing the parallel speedup to reduce. GraphMap does not suffer from such a synchronization bottleneck and hence does not experience a decrease in speedup. However, GraphMap's speedup saturates when the number of threads reaches a high number due to data sharing between threads.\par
\textbf{Summary.} According to the observations we make about GraphMap and Minimap, we conclude that storing minimizers instead of all \textit{k}-mers, as done by Minimap, does not affect the overall accuracy of the first three steps of the pipeline. Moreover, by storing minimizers, Minimap has a much lower memory usage and thus much higher performance than GraphMap.
\vspace{-10pt}
\subsection{Assembly Tools}\label{sec:assemblyresults}
As we discuss in Section~\ref{sec:assembly}, Canu and Miniasm are the commonly-used tools for this step.\footnote{In addition, we attempted to evaluate MECAT \cite{xiao2017mecat}, another assembler. We were unable to draw any meaningful conclusions from MECAT, as its memory usage exceeds the 1TB available in our \textit{big-mem} system.}
\vspace{-10pt}
\subsubsection*{Accuracy}
In order to compare the accuracy of these two tools, we categorize the results in Table~\ref{table:accuracyfirst3} based on assembly tools. We make the following observation.\par
\textbf{Observation 11: }\textit{Canu provides higher accuracy than Miniasm, with the help of the error-correction step that is present in its own pipeline.}
\vspace{-10pt}
\subsubsection*{Performance}
In order to compare the performance of Canu and Miniasm, we categorize the results in Table~\ref{table:perffirst3} based on assembly tools, in a way similar to what we did in Table~\ref{table:accuracyfirst3} for the accuracy analysis. We could not perform a scalability analysis for these tools since Canu has an auto-configuration mechanism for each sub-step of its own pipeline, which does not allow us to change the number of threads, and Miniasm does not support multi-threading. We make the following observation according to the results in Table~\ref{table:perffirst3}.\par
\textbf{Observation 12: }\textit{Canu is much more computationally intensive and greatly (\ie by 1096.3x) slower than Miniasm, because of its very expensive error-correction step.}\par
\textbf{Summary.} According to the observations we make about Canu and Miniasm, there is a tradeoff between accuracy and performance when deciding on the appropriate tool for this step. Canu produces highly accurate assemblies but it is resource intensive and slow. In contrast, Miniasm is a very fast assembler but it cannot produce as accurate draft assemblies as Canu. We suggest that Miniasm can potentially be used for fast initial analysis and then further polishing can be applied in the next step in order to produce higher-quality assemblies.
\vspace{-10pt}
\subsection{Read Mapping and Polishing Tools}\label{sec:polishresults}
\begin{table*}[b]
\caption{Accuracy analysis results for the full pipeline with a focus on the last two steps.}\label{table:accuracylast2}
\begin{tabularx}{\textwidth}{Y | r r r r r r}
\hline
& \specialcell{Number of\\Bases} & \specialcell{Number of\\Contigs} & \specialcell{Identity\\(\%)} & \specialcell{Coverage\\(\%)} & \specialcell{Number of\\Mismatches} & \specialcell{Number of\\Indels}\\ 
\hline
\begin{tabular}{p{0.25cm} | p{1.05cm} l p{1.05cm} l p{0.8cm} l p{1.4cm} r p{0.8cm}}1 & \textcolor{mypink}{Metrichor} & + & --- & + & \textcolor{mygreen}{Canu} & + &  \textcolor{myorange}{BWA-MEM} & + & \textcolor{mypurple}{Nanopolish}\end{tabular} & 4,683,072 & 1 & 99.48 & 99.93 & 8,198 & 15,581\\
\begin{tabular}{p{0.25cm} | p{1.05cm} l p{1.05cm} l p{0.8cm} l p{1.4cm} r p{0.8cm}}2 & \textcolor{mypink}{Metrichor} & + & \textcolor{myblue}{Minimap} & + & \textcolor{mygreen}{Miniasm} & + & \textcolor{myorange}{BWA-MEM} & + & \textcolor{mypurple}{Nanopolish}\end{tabular} & 4,540,352 & 1 & 92.33 & 96.31 & 162,884 & 182,965\\
\begin{tabular}{p{0.25cm} | p{1.05cm} l p{1.05cm} l p{0.8cm} l p{1.4cm} r p{0.8cm}}3 & \textcolor{mypink}{Metrichor} & + & \textcolor{myblue}{GraphMap} & + & \textcolor{mygreen}{Miniasm} & + & \textcolor{myorange}{BWA-MEM} & + & \textcolor{mypurple}{Nanopolish}\end{tabular} & 4,637,916 & 2 & 92.38 & 95.80 & 159,206 & 180,603\\
\hline
\hline
\begin{tabular}{p{0.25cm} | p{1.05cm} l p{1.05cm} l p{0.8cm} l p{1.4cm} r p{0.8cm}}4 & \textcolor{mypink}{Metrichor} & + & --- & + & \textcolor{mygreen}{Canu} & + &  \textcolor{myorange}{BWA-MEM} & + & \textcolor{mypurple}{Racon}\end{tabular} & 4,650,502 & 1 & 98.46 & 100.00 & 18,036 & 51,842\\
\begin{tabular}{p{0.25cm} | p{1.05cm} l p{1.05cm} l p{0.8cm} l p{1.4cm} r p{0.8cm}}5 & \textcolor{mypink}{Metrichor} & + & --- & + & \textcolor{mygreen}{Canu} & + & \textcolor{myorange}{Minimap} & + & \textcolor{mypurple}{Racon}\\\end{tabular} & 4,648,710 & 1 & 98.45 & 100.00 & 17,906 & 52,168\\
\begin{tabular}{p{0.25cm} | p{1.05cm} l p{1.05cm} l p{0.8cm} l p{1.4cm} r p{0.8cm}}6 & \textcolor{mypink}{Metrichor} & + & \textcolor{myblue}{Minimap} & + & \textcolor{mygreen}{Miniasm} & + & \textcolor{myorange}{BWA-MEM} & + & \textcolor{mypurple}{Racon}\end{tabular} & 4,598,267 & 1 & 97.70 & 99.91 & 24,014 & 82,906\\
\begin{tabular}{p{0.25cm} | p{1.05cm} l p{1.05cm} l p{0.8cm} l p{1.4cm} r p{0.8cm}}7 & \textcolor{mypink}{Metrichor} & + & \textcolor{myblue}{Minimap} & + & \textcolor{mygreen}{Miniasm} & + & \textcolor{myorange}{Minimap} & + & \textcolor{mypurple}{Racon}\end{tabular} & 4,600,109 & 1 & 97.78 & 100.00 & 23,339 & 79,721\\
\hline
\begin{tabular}{p{0.25cm} | p{1.05cm} l p{1.05cm} l p{0.8cm} l p{1.4cm} r p{0.8cm}}8 & \textcolor{mypink}{Nanonet} & + & --- & + & \textcolor{mygreen}{Canu} & + &  \textcolor{myorange}{BWA-MEM} & + & \textcolor{mypurple}{Racon}\end{tabular} & 4,622,285 & 1 & 98.48 & 100.00 & 16,872 & 52,509\\
\begin{tabular}{p{0.25cm} | p{1.05cm} l p{1.05cm} l p{0.8cm} l p{1.4cm} r p{0.8cm}}9 & \textcolor{mypink}{Nanonet} & + & --- & + & \textcolor{mygreen}{Canu} & + & \textcolor{myorange}{Minimap} & + & \textcolor{mypurple}{Racon}\\\end{tabular} & 4,620,597 & 1 & 98.49 & 100.00 & 16,874 & 52,232\\
\begin{tabular}{p{0.25cm} | p{1.05cm} l p{1.05cm} l p{0.8cm} l p{1.4cm} r p{0.8cm}}10 & \textcolor{mypink}{Nanonet} & + & \textcolor{myblue}{Minimap} & + & \textcolor{mygreen}{Miniasm} & + & \textcolor{myorange}{BWA-MEM} & + & \textcolor{mypurple}{Racon}\end{tabular} & 4,593,402 & 1 & 98.01 & 99.97 & 20,322 & 72,284\\
\begin{tabular}{p{0.25cm} | p{1.05cm} l p{1.05cm} l p{0.8cm} l p{1.4cm} r p{0.8cm}}11 & \textcolor{mypink}{Nanonet} & + & \textcolor{myblue}{Minimap} & + & \textcolor{mygreen}{Miniasm} & + & \textcolor{myorange}{Minimap} & + & \textcolor{mypurple}{Racon}\end{tabular} & 4,592,907 & 1 & 98.04 & 100.00 & 20,170 & 70,705\\
\hline
\begin{tabular}{p{0.25cm} | p{1.05cm} l p{1.05cm} l p{0.8cm} l p{1.4cm} r p{0.8cm}}12 & \textcolor{mypink}{Scrappie} & + & --- & + & \textcolor{mygreen}{Canu} & + &  \textcolor{myorange}{BWA-MEM} & + & \textcolor{mypurple}{Racon}\end{tabular} & 4,673,871 & 1 & 98.40 & 99.98 & 13,583 & 60,612\\
\begin{tabular}{p{0.25cm} | p{1.05cm} l p{1.05cm} l p{0.8cm} l p{1.4cm} r p{0.8cm}}13 & \textcolor{mypink}{Scrappie} & + & --- & + &\textcolor{mygreen}{Canu} & + & \textcolor{myorange}{Minimap} & + & \textcolor{mypurple}{Racon}\\\end{tabular} & 4,673,606 & 1 & 98.40 & 99.98 & 13,798 & 60,423\\
\begin{tabular}{p{0.25cm} | p{1.05cm} l p{1.05cm} l p{0.8cm} l p{1.4cm} r p{0.8cm}}14 & \textcolor{mypink}{Scrappie} & + & \textcolor{myblue}{Minimap} & + & \textcolor{mygreen}{Miniasm} & + & \textcolor{myorange}{BWA-MEM} & + & \textcolor{mypurple}{Racon}\end{tabular} & 5,157,041 & 8 & 97.87 & 99.80 & 18,085 & 78,492\\
\begin{tabular}{p{0.25cm} | p{1.05cm} l p{1.05cm} l p{0.8cm} l p{1.4cm} r p{0.8cm}}15 & \textcolor{mypink}{Scrappie} & + & \textcolor{myblue}{Minimap} & + & \textcolor{mygreen}{Miniasm} & + & \textcolor{myorange}{Minimap} & + &\textcolor{mypurple}{Racon}\end{tabular} & 5,156,375 & 8 & 97.87 & 99.94 & 17,922 & 77,807\\
\hline
\begin{tabular}{p{0.25cm} | p{1.05cm} l p{1.05cm} l p{0.8cm} l p{1.4cm} r p{0.8cm}}16 & \textcolor{mypink}{Nanocall} & + & --- & + & \textcolor{mygreen}{Canu} & + &  \textcolor{myorange}{BWA-MEM} & + & \textcolor{mypurple}{Racon}\end{tabular} & 1,383,851 & 86 & 93.49 & 28.82 & 19,057 & 65,244\\
\begin{tabular}{p{0.25cm} | p{1.05cm} l p{1.05cm} l p{0.8cm} l p{1.4cm} r p{0.8cm}}17 & \textcolor{mypink}{Nanocall} & + & --- & + & \textcolor{mygreen}{Canu} & + & \textcolor{myorange}{Minimap} & + & \textcolor{mypurple}{Racon}\\\end{tabular} & 1,367,834 & 86 & 94.43 & 28.74 & 15,610 & 55,275\\
\begin{tabular}{p{0.25cm} | p{1.05cm} l p{1.05cm} l p{0.8cm} l p{1.4cm} r p{0.8cm}}18 & \textcolor{mypink}{Nanocall} & + & \textcolor{myblue}{Minimap} & + & \textcolor{mygreen}{Miniasm} & + & \textcolor{myorange}{BWA-MEM} & + & \textcolor{mypurple}{Racon}\end{tabular} & 4,707,961 & 5 & 90.75 & 97.11 & 91,502 & 347,005\\
\begin{tabular}{p{0.25cm} | p{1.05cm} l p{1.05cm} l p{0.8cm} l p{1.4cm} r p{0.8cm}}19 & \textcolor{mypink}{Nanocall} & + & \textcolor{myblue}{Minimap} & + & \textcolor{mygreen}{Miniasm} & + & \textcolor{myorange}{Minimap} & + & \textcolor{mypurple}{Racon}\end{tabular} & 4,673,069 & 5 & 92.23 & 97.10 & 72,646 & 291,918\\
\hline
\begin{tabular}{p{0.25cm} | p{1.05cm} l p{1.05cm} l p{0.8cm} l p{1.4cm} r p{0.8cm}}20 & \textcolor{mypink}{DeepNano} & + & --- & + & \textcolor{mygreen}{Canu} & + &  \textcolor{myorange}{BWA-MEM} & + & \textcolor{mypurple}{Racon}\end{tabular} & 7,429,290 & 106 & 96.46 & 99.24 & 27,811 & 102,682\\
\begin{tabular}{p{0.25cm} | p{1.05cm} l p{1.05cm} l p{0.8cm} l p{1.4cm} r p{0.8cm}}21 & \textcolor{mypink}{DeepNano} & + & --- & + & \textcolor{mygreen}{Canu} & + & \textcolor{myorange}{Minimap} & + & \textcolor{mypurple}{Racon}\\\end{tabular} & 7,404,454 & 106 & 96.03 & 99.21 & 34,023 & 110,640\\
\begin{tabular}{p{0.25cm} | p{1.05cm} l p{1.05cm} l p{0.8cm} l p{1.4cm} r p{0.8cm}}22 & \textcolor{mypink}{DeepNano} & + & \textcolor{myblue}{Minimap} & + & \textcolor{mygreen}{Miniasm} & + & \textcolor{myorange}{BWA-MEM} & + & \textcolor{mypurple}{Racon}\end{tabular} & 4,566,253 & 1 & 96.76 & 99.86 & 25,791 & 125,386\\
\begin{tabular}{p{0.25cm} | p{1.05cm} l p{1.05cm} l p{0.8cm} l p{1.4cm} r p{0.8cm}}23 & \textcolor{mypink}{DeepNano} & + & \textcolor{myblue}{Minimap} & + & \textcolor{mygreen}{Miniasm} & + & \textcolor{myorange}{Minimap} & + & \textcolor{mypurple}{Racon}\end{tabular} & 4,571,810 & 1 & 96.90 & 99.97 & 24,994 & 119,519\\
\hline
\end{tabularx}
\end{table*}
\begin{table*}[t]
\vspace{-5pt}
\caption{Performance analysis results for the full pipeline with a focus on the last two steps.}\label{table:perflast2}
\begin{tabularx}{\textwidth}{Y | r r r | r r r}
\hline
& \multicolumn{3}{c|}{\textcolor{myorange}{\specialcellll{\textbf{Step 4:} Read Mapper}}} &  \multicolumn{3}{c}{\textcolor{mypurple}{\specialcellll{\textbf{Step 5:} Polisher}}}\\
\cline{2-7}
& \specialcell{Wall\\Clock\\Time\\(h:m:s)} & \specialcell{CPU Time\\(h:m:s)} & \specialcell{Memory\\Usage\\(GB)} & \specialcell{Wall\\Clock\\Time\\(h:m:s)} & \specialcell{CPU Time\\(h:m:s)} & \specialcell{Memory\\Usage\\(GB)}\\ 
\hline
\begin{tabular}{p{0.3cm} | p{1.2cm} l p{1.1cm} l p{0.9cm} l p{1.4cm} r p{0.8cm}}1 & \textcolor{mypink}{Metrichor} & + & --- & + & \textcolor{mygreen}{Canu} & + &  \textcolor{myorange}{BWA-MEM} & + & \textcolor{mypurple}{Nanopolish}\end{tabular} & 24:43 & 15:47:21 & 5.26 & 5:51:00 & 191:18:52 & 13.38\\
\begin{tabular}{p{0.3cm} | p{1.2cm} l p{1.1cm} l p{0.9cm} l p{1.4cm} r p{0.8cm}}2 & \textcolor{mypink}{Metrichor} & + & \textcolor{myblue}{Minimap} & + & \textcolor{mygreen}{Miniasm} & + & \textcolor{myorange}{BWA-MEM} & + & \textcolor{mypurple}{Nanopolish}\end{tabular} & 12:33 & 7:50:54 & 3.75 & 122:52:00 & 4458:36:10 & 31.36\\
\begin{tabular}{p{0.3cm} | p{1.2cm} l p{1.1cm} l p{0.9cm} l p{1.4cm} r p{0.8cm}}3 & \textcolor{mypink}{Metrichor} & + & \textcolor{myblue}{GraphMap} & + & \textcolor{mygreen}{Miniasm} & + & \textcolor{myorange}{BWA-MEM} & + & \textcolor{mypurple}{Nanopolish}\end{tabular} & 12:47 & 7:57:58 & 3.60 & 129:46:00 & 4799:03:51 & 31.31\\
\hline
\hline
\begin{tabular}{p{0.3cm} | p{1.2cm} l p{1.1cm} l p{0.9cm} l p{1.4cm} r p{0.8cm}}4 & \textcolor{mypink}{Metrichor} & + & --- & + & \textcolor{mygreen}{Canu} & + &  \textcolor{myorange}{BWA-MEM} & + & \textcolor{mypurple}{Racon}\end{tabular} & 24:20 & 15:43:40 & 6.60 & 14:44 & 9:09:22 & 8.11\\
\begin{tabular}{p{0.3cm} | p{1.2cm} l p{1.1cm} l p{0.9cm} l p{1.4cm} r p{0.8cm}}5 & \textcolor{mypink}{Metrichor} & + & --- & + & \textcolor{mygreen}{Canu} & + & \textcolor{myorange}{Minimap} & + & \textcolor{mypurple}{Racon}\end{tabular} & 3 & 1:35 & 0.26 & 15:12 & 9:45:33 & 14.55\\
\begin{tabular}{p{0.3cm} | p{1.2cm} l p{1.1cm} l p{0.9cm} l p{1.4cm} r p{0.8cm}}6 & \textcolor{mypink}{Metrichor} & + & \textcolor{myblue}{Minimap} & + & \textcolor{mygreen}{Miniasm} & + & \textcolor{myorange}{BWA-MEM} & + & \textcolor{mypurple}{Racon}\end{tabular} & 12:10 & 7:48:10 & 5.19 & 15:43 & 9:33:39 & 9.98\\
\begin{tabular}{p{0.3cm} | p{1.2cm} l p{1.1cm} l p{0.9cm} l p{1.4cm} r p{0.8cm}}7 & \textcolor{mypink}{Metrichor} & + & \textcolor{myblue}{Minimap} & + & \textcolor{mygreen}{Miniasm} & + & \textcolor{myorange}{Minimap} & + & \textcolor{mypurple}{Racon}\end{tabular} & 3 & 1:24 & 0.26 & 20:28 & 8:57:40 & 18.24\\
\hline
\begin{tabular}{p{0.3cm} | p{1.2cm} l p{1.1cm} l p{0.9cm} l p{1.4cm} r p{0.8cm}}8 & \textcolor{mypink}{Nanonet} & + & --- & + & \textcolor{mygreen}{Canu} & + &  \textcolor{myorange}{BWA-MEM} & + & \textcolor{mypurple}{Racon}\end{tabular} & 9:08 & 5:53:18 & 4.84 & 6:33 & 4:02:10 & 4.47\\
\begin{tabular}{p{0.3cm} | p{1.2cm} l p{1.1cm} l p{0.9cm} l p{1.4cm} r p{0.8cm}}9 & \textcolor{mypink}{Nanonet} & + & --- & + & \textcolor{mygreen}{Canu} & + & \textcolor{myorange}{Minimap} & + & \textcolor{mypurple}{Racon}\end{tabular} & 2 & 54 & 0.26 & 6:45 & 4:17:26 & 7.93\\
\begin{tabular}{p{0.3cm} | p{1.2cm} l p{1.1cm} l p{0.9cm} l p{1.4cm} r p{0.8cm}}10 & \textcolor{mypink}{Nanonet} & + & \textcolor{myblue}{Minimap} & + & \textcolor{mygreen}{Miniasm} & + & \textcolor{myorange}{BWA-MEM} & + & \textcolor{mypurple}{Racon}\end{tabular} & 4:40 & 2:58:02 & 3.88 & 7:08 & 4:19:30 & 5.35\\
\begin{tabular}{p{0.3cm} | p{1.2cm} l p{1.1cm} l p{0.9cm} l p{1.4cm} r p{0.8cm}}11 & \textcolor{mypink}{Nanonet} & + & \textcolor{myblue}{Minimap} & + & \textcolor{mygreen}{Miniasm} & + & \textcolor{myorange}{Minimap} & + & \textcolor{mypurple}{Racon}\end{tabular} & 2 & 46 & 0.26 & 7:01 & 4:18:48 & 9.53\\
\hline
\begin{tabular}{p{0.3cm} | p{1.2cm} l p{1.1cm} l p{0.9cm} l p{1.4cm} r p{0.8cm}}12 & \textcolor{mypink}{Scrappie} & + & --- & + & \textcolor{mygreen}{Canu} & + &  \textcolor{myorange}{BWA-MEM} & + & \textcolor{mypurple}{Racon}\end{tabular} & 33:41 & 21:11:06 & 8.66 & 13:32 & 8:24:44 & 7.58\\
\begin{tabular}{p{0.3cm} | p{1.2cm} l p{1.1cm} l p{0.9cm} l p{1.4cm} r p{0.8cm}}13 & \textcolor{mypink}{Scrappie} & + & --- & + & \textcolor{mygreen}{Canu} & + & \textcolor{myorange}{Minimap} & + & \textcolor{mypurple}{Racon}\end{tabular} & 3 & 1:39 & 0.27 & 18:45 & 7:43:17 & 13.20\\
\begin{tabular}{p{0.3cm} | p{1.2cm} l p{1.1cm} l p{0.9cm} l p{1.4cm} r p{0.8cm}}14 & \textcolor{mypink}{Scrappie} & + & \textcolor{myblue}{Minimap} & + & \textcolor{mygreen}{Miniasm} & + & \textcolor{myorange}{BWA-MEM} & + & \textcolor{mypurple}{Racon}\end{tabular} & 22:41 & 14:31:00 & 6.08 & 14:37 & 8:53:59 & 9.50\\
\begin{tabular}{p{0.3cm} | p{1.2cm} l p{1.1cm} l p{0.9cm} l p{1.4cm} r p{0.8cm}}15 & \textcolor{mypink}{Scrappie} & + & \textcolor{myblue}{Minimap} & + & \textcolor{mygreen}{Miniasm} & + & \textcolor{myorange}{Minimap} & + & \textcolor{mypurple}{Racon}\end{tabular} & 3 & 1:27 & 0.27 & 15:10 & 9:02:45 & 12.72\\
\hline
\begin{tabular}{p{0.3cm} | p{1.2cm} l p{1.1cm} l p{0.9cm} l p{1.4cm} r p{0.8cm}}16 & \textcolor{mypink}{Nanocall} & + & --- & + & \textcolor{mygreen}{Canu} & + &  \textcolor{myorange}{BWA-MEM} & + & \textcolor{mypurple}{Racon}\end{tabular} & 4:52 & 3:01:15 & 3.80 & 11:07 & 3:26:52 & 5.63\\
\begin{tabular}{p{0.3cm} | p{1.2cm} l p{1.1cm} l p{0.9cm} l p{1.4cm} r p{0.8cm}}17 & \textcolor{mypink}{Nanocall} & + & --- & + & \textcolor{mygreen}{Canu} & + & \textcolor{myorange}{Minimap} & + & \textcolor{mypurple}{Racon}\end{tabular} & 3 & 1:16 & 0.22 & 7:28 & 2:50:35 & 3.62\\
\begin{tabular}{p{0.3cm} | p{1.2cm} l p{1.1cm} l p{0.9cm} l p{1.4cm} r p{0.8cm}}18 & \textcolor{mypink}{Nanocall} & + & \textcolor{myblue}{Minimap} & + & \textcolor{mygreen}{Miniasm} & + & \textcolor{myorange}{BWA-MEM} & + & \textcolor{mypurple}{Racon}\end{tabular} & 16:06 & 10:27:20 & 5.06 & 18:56 & 11:32:45 & 11.47\\
\begin{tabular}{p{0.3cm} | p{1.2cm} l p{1.1cm} l p{0.9cm} l p{1.4cm} r p{0.8cm}}19 & \textcolor{mypink}{Nanocall} & + & \textcolor{myblue}{Minimap} & + & \textcolor{mygreen}{Miniasm} & + & \textcolor{myorange}{Minimap} & + & \textcolor{mypurple}{Racon}\end{tabular} & 4 & 1:18 & 0.26 & 11:49 & 7:08:59 & 10.98\\
\hline
\begin{tabular}{p{0.3cm} | p{1.2cm} l p{1.1cm} l p{0.9cm} l p{1.4cm} r p{0.8cm}}20 & \textcolor{mypink}{DeepNano} & + & --- & + & \textcolor{mygreen}{Canu} & + &  \textcolor{myorange}{BWA-MEM} & + & \textcolor{mypurple}{Racon}\end{tabular} & 17:36 & 11:30:20 & 4.43 & 12:48 & 7:13:04 & 8.88\\
\begin{tabular}{p{0.3cm} | p{1.2cm} l p{1.1cm} l p{0.9cm} l p{1.4cm} r p{0.8cm}}21 & \textcolor{mypink}{DeepNano} & + & --- & + & \textcolor{mygreen}{Canu} & + & \textcolor{myorange}{Minimap} & + & \textcolor{mypurple}{Racon}\end{tabular} & 3 & 1:24 & 0.28 & 11:39 & 6:55:01 & 3.73\\
\begin{tabular}{p{0.3cm} | p{1.2cm} l p{1.1cm} l p{0.9cm} l p{1.4cm} r p{0.8cm}}22 & \textcolor{mypink}{DeepNano} & + & \textcolor{myblue}{Minimap} & + & \textcolor{mygreen}{Miniasm} & + & \textcolor{myorange}{BWA-MEM} & + & \textcolor{mypurple}{Racon}\end{tabular} & 8:15 & 5:22:29 & 4.11 & 14:16 & 8:34:32 & 10.30\\
\begin{tabular}{p{0.3cm} | p{1.2cm} l p{1.1cm} l p{0.9cm} l p{1.4cm} r p{0.8cm}}23 & \textcolor{mypink}{DeepNano} & + & \textcolor{myblue}{Minimap} & + & \textcolor{mygreen}{Miniasm} & + & \textcolor{myorange}{Minimap} & + & \textcolor{mypurple}{Racon}\end{tabular} & 3 & 1:10 & 0.26 & \textcolor{white}{xxx:}12:29 & \textcolor{white}{xxx}7:55:32 & 17.11\\
\hline
\end{tabularx}
\vspace{-15pt}
\end{table*}
As we discuss in Section~\ref{sec:polish}, further polishing may be required for improving the accuracy of the low-quality draft assemblies. For this purpose, after aligning the reads to the generated draft assembly with BWA-MEM or Minimap,\footnote{We do not discuss these tools in great detail here since they perform read mapping, which is commonly analyzed and relatively well understood (\eg see \cite{li2009fast,langmead2009ultrafast,alkan2009personalized,hach2010mrsfast,schatz2009cloudburst,li2008mapping,kim2017grim,xin2015shifted,alser2017gatekeeper,alser2017magnet,weese2009razers,lee2014mosaik,rumble2009shrimp,david2011shrimp2,hatem2013benchmarking,olson2012hardware,fonseca2012tools,li2010fast,siragusa2013fast}).}\textsuperscript{,}\footnote{Minimap2 \cite{li2017minimap2} is a recently-released successor to Minimap. We compare Minimap2 to BWA-MEM and to Minimap, and make two observations. First, Minimap2 significantly outperforms BWA-MEM. Since Minimap2 can produce SAM alignments (which BWA-MEM produces), we can replace BWA-MEM with Minimap2 in future genome assembly pipelines. Second, Minimap2 has similar accuracy and performance compared to Minimap. This is because Minimap2 and Minimap employ similar indexing and seeding algorithms \cite{li2017minimap2}, and the new features of Minimap2 (more accurate chaining, base-level alignment, support for spliced alignment) are \emph{not} used in the pipeline we analyze. As a result, our findings for Minimap generally remain the same for Minimap2.} one can use Nanopolish or Racon to perform polishing and obtain improved assemblies (\ie consensus sequences).\par
Nanopolish accepts mappings only in \textit{Sequence Alignment/Map (SAM)} format \cite{li2009sequence} and it works only with draft assemblies generated with the Metrichor-basecalled reads. On the other hand, Racon accepts both \textit{Pairwise Mapping format (PAF)} mappings \cite{li2016minimap} and \textit{SAM}-format mappings, but it requires the input reads and draft assembly files to be in \textit{fastq} format \cite{cock2009sanger}, which includes quality scores. However, by using the \texttt{-bq -1} parameter, it is possible to disable the filtering used in Racon, which requires quality scores. Since our basecalled reads are in \textit{fasta} format \cite{pearson1988improved}, in our experiments, we convert these \textit{fasta} files into \textit{fastq} files and disable the filtering with the corresponding parameter.\par
BWA-MEM generates mappings in \textit{SAM} format whereas Minimap generates mappings in \textit{PAF} format. Since Nanopolish requires \textit{SAM} format input, we generate the mappings only with BWA-MEM and use them for Nanopolish polishing, in our analysis. On the other hand, since Racon accepts both formats, we generate the mappings and the overlaps with both BWA-MEM and Minimap, respectively, and use them for Racon polishing, in our analysis.
\vspace{-10pt}
\begin{figure*}[bh!]
\vspace{-15pt}
\includegraphics[width=\textwidth,height=11.8cm]{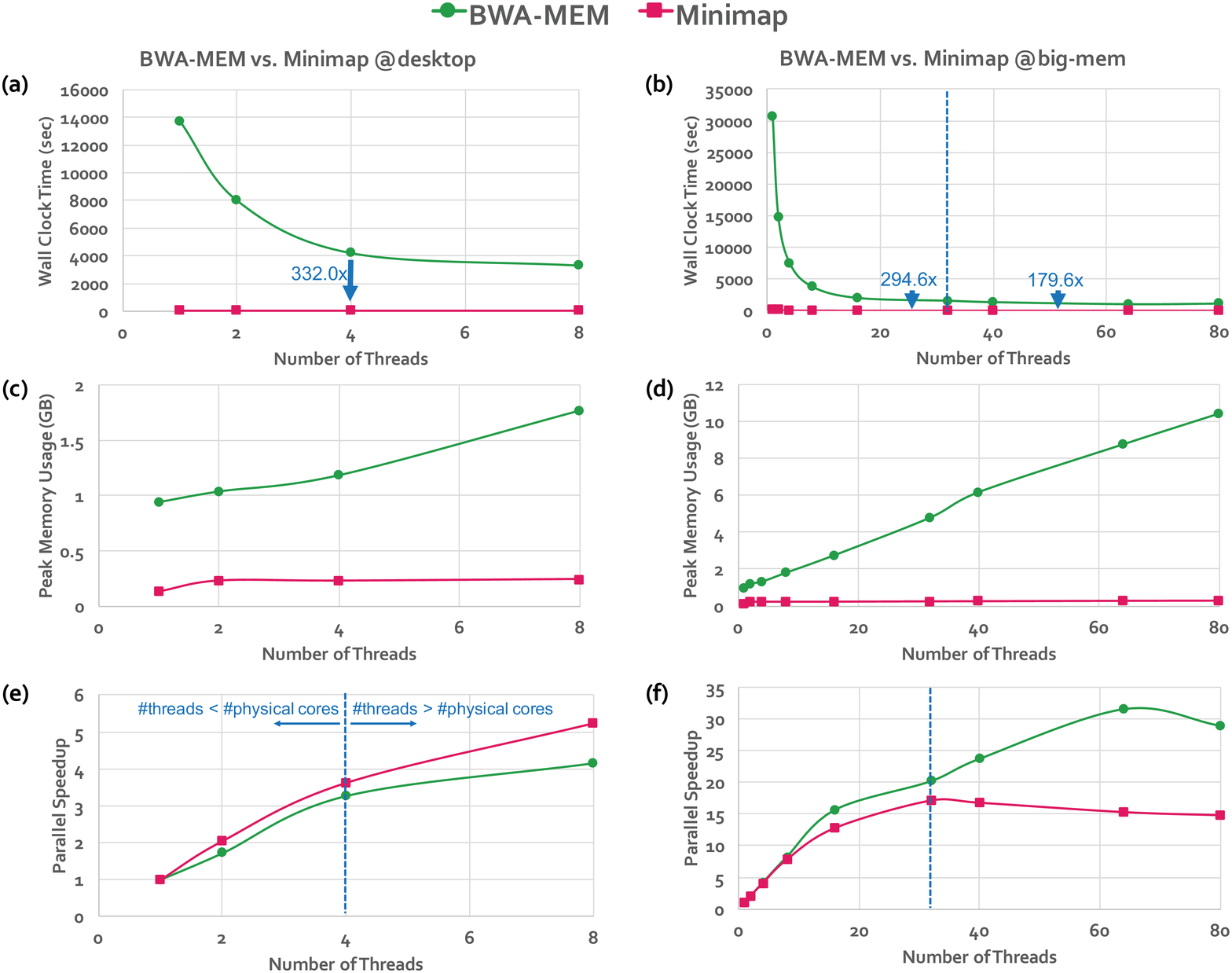}
\caption{Scalability results of BWA-MEM and Minimap.}\label{fig:mapperplot}
Wall clock time (a, b), peak memory usage (c, d), and parallel speedup (e, f) results obtained on the \textit{desktop} and \textit{big-mem} systems. The left column (a, c, e) shows the results from the \textit{desktop} system and the right column (b, d, f) shows the results from the \textit{big-mem} system.
\vspace{-10pt}
\end{figure*}
\vspace{-5pt}
\subsubsection*{Accuracy}
Table~\ref{table:accuracylast2} presents the accuracy metrics results for Nanopolish (\ie Rows 1-3) and Racon (\ie Rows 4-23) pipelines. Based on these results, we make two observations.\par
\textbf{Observation 13: }\textit{Both Nanopolish and Racon significantly increase the accuracy of the draft assemblies.}\par
For example, Nanopolish increases the identity and coverage of the draft assembly generated with the Metrichor+Minimap+Miniasm pipeline from 87.71\% and 94.85\% (Row 2 of Table~\ref{table:accuracyfirst3}), respectively, to 92.33\% and 96.31\% (Row 2 of Table~\ref{table:accuracylast2}). Similarly, Racon increases them to 97.70\% and 99.91\% (Rows 6--7 of Table~\ref{table:accuracylast2}), respectively.\par
\textbf{Observation 14: }\textit{For Racon, the choice of read mapper does not affect the accuracy of the polishing step.}\par
We observe that using BWA-MEM or Minimap to generate the mappings for Racon results in almost identical accuracy metric results. For example, when we use BWA-MEM before Racon for the draft assembly generated with the Metrichor + Canu pipeline (Row 4 of Table~\ref{table:accuracylast2}), Racon results with 98.46\% identity, 100.00\% coverage, 18,036 mismatches and 51,842 indels. When we use Minimap, instead (Row 5 of Table~\ref{table:accuracylast2}), Racon results with 98.45\% identity, 100.00\% coverage, 17,096 mismatches and 52,168 indels, which is almost identical to the BWA-MEM results.  
\vspace{-5pt}
\subsubsection*{Performance}
In the first part of the performance analysis for Nanopolish, we divide the draft assemblies into 50kb-segments and polish 4 of these segments in parallel with 10 threads for each segment. For Racon, each draft assembly is polished using 40 threads, but the tool, by default, divides the input sequence into windows of 20kb length. Table~\ref{table:perflast2} presents the performance results for Nanopolish (\ie Rows 1-3) and Racon (\ie Rows 4-23) pipelines. Based on these results, we make the following two observations.\par
\textbf{Observation 15: }\textit{Nanopolish is computationally much more intensive and thus greatly slower than Racon.}\par
Nanopolish runs take days to complete whereas Racon runs take minutes. This is mainly because Nanopolish works on each base individually, whereas Racon works on the windows. Since each window is much longer (\ie 20kb) than a single base, the computational workload is greatly smaller in Racon. Also, Racon only uses the mappings/overlaps for polishing, whereas Nanopolish uses raw signal data and an HMM-based approach in order to generate the consensus sequence, which is computationally more expensive.\par
\textbf{Observation 16: }\textit{BWA-MEM is computationally more expensive than Minimap.}\par
Although the choice of BWA-MEM and Minimap for the read mapping step does not affect the accuracy of the polishing step, these two tools have a significant difference in performance.\par
For a deeper performance analysis of these read mapping and polishing tools, we perform a scalability analysis for each read mapper and each polisher by running them on the \textit{desktop} system and the \textit{big-mem} system separately, with 1, 2, 4, 8 (maximum for \textit{desktop} server), 16, 32, 40, 64 and 80 (maximum for \textit{big-mem} server) threads, and measuring the performance metrics. Figure~\ref{fig:mapperplot} shows the the speed, memory usage and parallel speedup of BWA-MEM and Minimap. We make two observations.\par
\textbf{Observation 17: }\textit{On both systems, Minimap is greatly faster than BWA-MEM (cf. Figure~\ref{fig:mapperplot}a and~\ref{fig:mapperplot}b). However, when the number of threads reaches high value, Minimap's performance degrades due to the synchronization overhead between its threads (cf. Figure~\ref{fig:mapperplot}f).}\par
On the \textit{desktop} system, Minimap is 332.0x faster than BWA-MEM, on average (see Figure~\ref{fig:mapperplot}a). On the \textit{big-mem} system, Minimap is 294.6x and 179.6x faster than BWA-MEM, on average, when the number of threads is smaller and greater than 32, respectively. This is due to the synchronization overhead that increases with the number of threads used in Minimap (see Observation 10). As we also show in Figure~\ref{fig:mapperplot}f, Minimap's speedup reduces when the number of threads exceeds 32, which is another indication of the synchronization overhead that causes Minimap to slow down.\par
\textbf{Observation 18: }\textit{Minimap's memory usage is independent of the number of threads and stays constant. In contrast, BWA-MEM's memory usage increases linearly with the number of threads (cf. Figure~\ref{fig:mapperplot}c and~\ref{fig:mapperplot}d).}\par
In Minimap, memory usage is dependent on the hash table size and is independent of number of threads (see Observation 8). In contrast, in BWA-MEM, each thread separately performs computation for different groups of reads (as in Scrappie and Nanocall, see Observation 4). This causes the linear increase in memory usage of BWA-MEM when the number of threads increases.\par
\begin{figure}[H]
\includegraphics[width=\columnwidth,height=12.5cm]{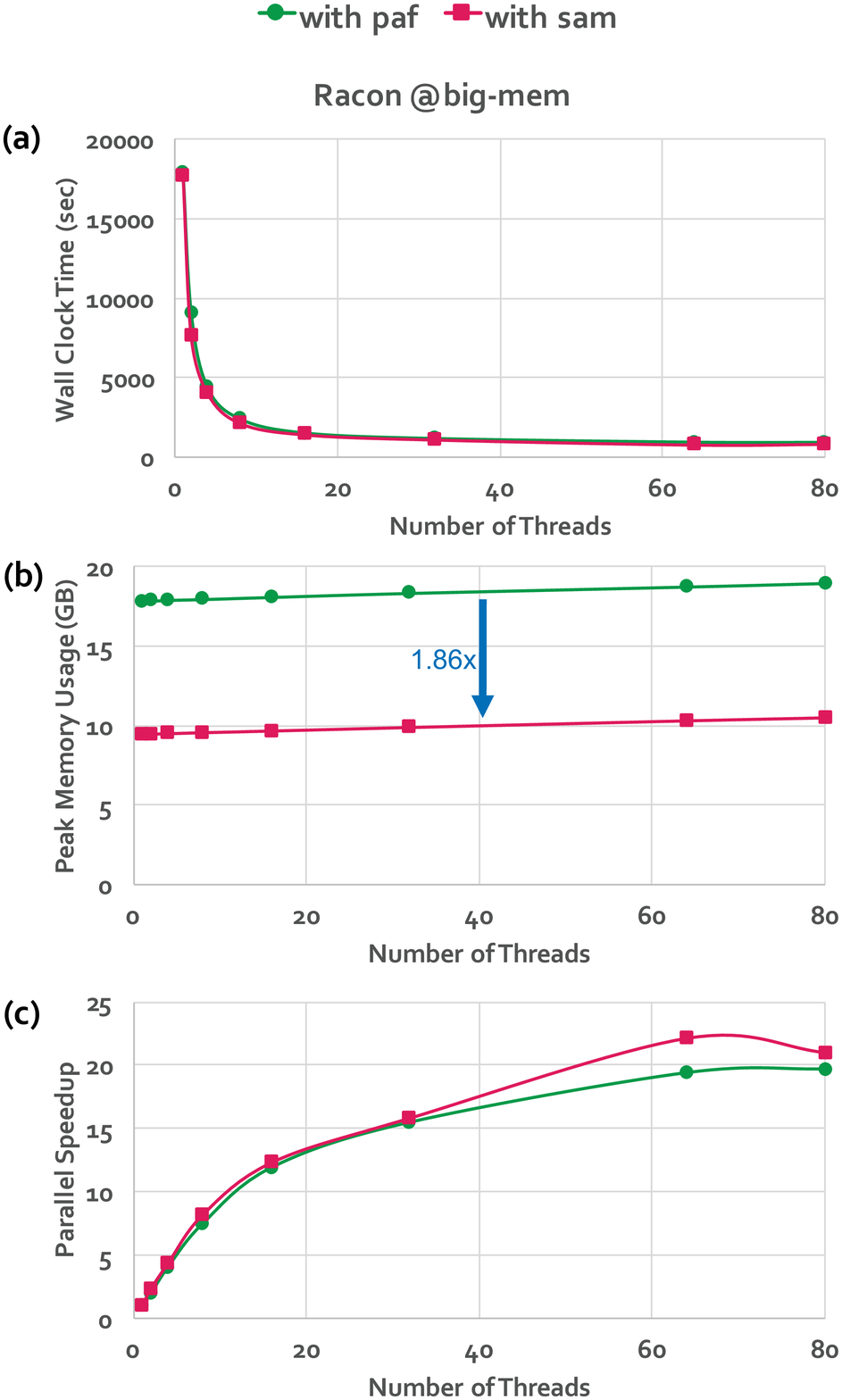}
\caption{Scalability results of Racon.}\label{fig:raconplot}
Wall clock time (a), peak memory usage (b), and parallel speedup (c) results obtained on the \textit{big-mem} system.
\vspace{-10pt}
\end{figure}

Figure~\ref{fig:raconplot} shows the scalability results for Racon on the \textit{big-mem} system. We obtain the results on both of the systems. However, we only show the results for the \textit{big-mem} system since the results for both of the systems are similar. We separately test the tool by using \textit{PAF} mappings and \textit{SAM} mappings. Based on the results, we make the following observation.\par 
\textbf{Observation 19: }\textit{Racon's memory usage is independent of the number of threads for both \textit{PAF} mode and \textit{SAM} mode. However, the memory usage of \textit{PAF} mode is 1.86x higher than the memory usage of \textit{SAM} mode, on average (cf. Figure~\ref{fig:raconplot}b).}\par
The memory usage of Racon depends on the number of mappings received from the fourth step since Racon performs polishing by using these mappings. Racon's memory usage is higher for the PAF mode because the number of mappings stored in the PAF files is greater than the number of mappings stored in the SAM files (\ie 1.4x). However, using \textit{PAF} mappings or \textit{SAM} mappings do not significantly affect the speed (see Figure~\ref{fig:raconplot}a) and the parallel speedup (see Figure~\ref{fig:raconplot}c) of Racon.\par
\begin{figure}
\includegraphics[width=\columnwidth,height=12.5cm]{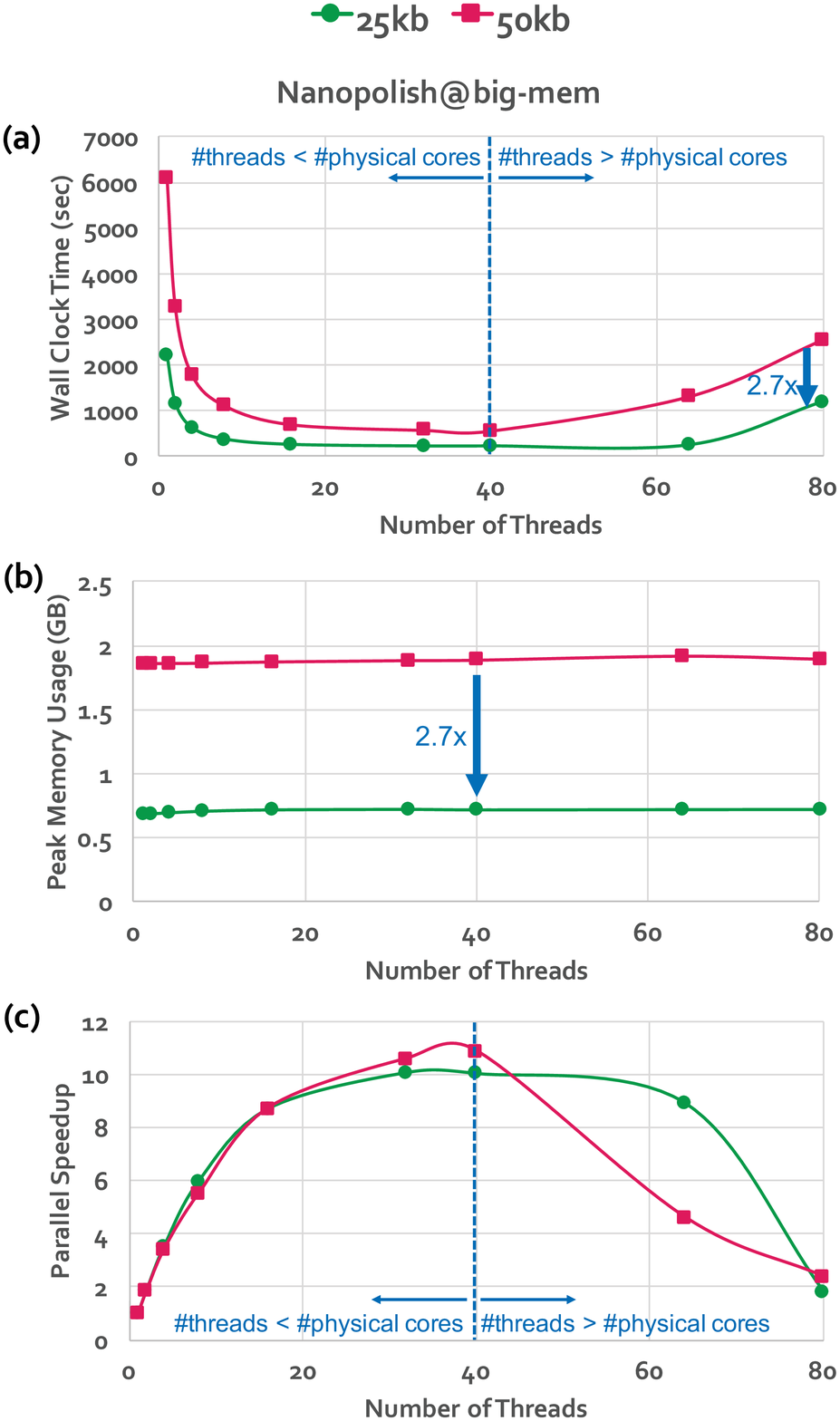}
\caption{Scalability results of Nanopolish.}\label{fig:nanopolishplot}
Wall clock time (a), peak memory usage (b), and parallel speedup (c) results obtained on the \textit{big-mem} system.
\vspace{-10pt}
\end{figure}

Figure~\ref{fig:nanopolishplot} shows the scalability results for Nanopolish. We test the tool by separately using a 25kb and a 50kb segment length to assess the scalability of the tool with respect to the segment length, in addition to the scalability with respect to the number of threads. We measure the performance metrics. We only show the results for the \textit{big-mem} system since the results for both of the systems are similar. Based on the results, we make the following observation.\par 
\textbf{Observation 20: }\textit{Nanopolish's memory usage is independent of the number of threads. However, its memory usage in dependent on the segment length (cf. Figure~\ref{fig:nanopolishplot}b).}\par
The memory usage of Nanopolish is not affected by the number of threads. However, it is dependent on the segment length. Nanopolish uses more memory for longer segments. When the segment length is doubled from 25kb to 50kb, the increase in the memory usage (\ie 2.7x) is greater than 2.0x. This is because the memory usage of Nanopolish depends both on the length of the segment and the number of read mappings that map to this segment. For both of the segments, the memory usage also affects the speed. The Nanopolish run for the 25kb-segment is 2.7x faster than the run for the 50kb-segment (see Figure~\ref{fig:nanopolishplot}a).\par
\textbf{Observation 21: }\textit{Nanopolish's performance greatly degrades when the number of threads exceeds the number of physical cores (cf. Figure~\ref{fig:nanopolishplot}c).}\par
Hyper-threading causes a slowdown for Nanopolish because of the CPU-intensive workload of Nanopolish and the resulting high contention in the shared resources between the threads executing on the same core, as we discuss in Observation 5.\par
\textbf{Summary.} Based on the observations we make about tools for the optional last two steps of the pipeline, we conclude that further polishing can significantly increase the accuracy of the assemblies. Since BWA-MEM and Nanopolish are more resource-intensive than Minimap and Racon, pipelines with Minimap and Racon can provide a significant speedup compared to the pipelines with BWA-MEM and Nanopolish, while resulting with high-quality consensus sequences. 

\vspace{-15pt}
\section{Recommendations} \label{sec:recommendations}
\subsection*{Recommendations for Tool Users}
Based on the results we have collected and observations we have made for each step of the genome assembly pipeline using nanopore sequence data and the associated tools, we make the following major recommendations for the current and future tool users.
\begin{itemize}
\item ONT's basecalling tools, Metrichor, Nanonet, and Scrappie, are the best choices for the basecalling step in terms of both accuracy and performance. Among these tools, Scrappie is the newest, fastest and most accurate basecaller. Thus, we recommend using Scrappie for the basecalling step (See analysis in Section~\ref{sec:basecallresults}).
\item For the read-to-read overlap finding step, Minimap is faster than GraphMap, and it requires low memory. Also, it has similar accuracy to GraphMap. Thus, we recommend Minimap for the read-to-read overlap finding step (See analysis in Section~\ref{sec:overlapresults}).
\item For the assembly step, if execution time is not an important concern, we recommend using Canu since it produces much more accurate assemblies. However, for a fast initial analysis, we recommend using Miniasm since it is fast and its accuracy can be increased with an additional polishing step. If Miniasm is used for assembly, we definitely recommend further polishing to increase the accuracy of the final assembly (See analysis in Section~\ref{sec:assemblyresults}). Even though polishing takes a similar amount of time if we use Miniasm or Canu, the accuracy improvements are much \emph{smaller} for a genome assembled using Canu. We hope that future work can improve the performance of polishing when the assembled genome already has high accuracy, to reduce the execution time of the overall assembly pipeline.
\item For the polishing step, we recommend using Racon since it is much faster than Nanopolish. Racon also produces highly-accurate assemblies (See analysis in Section~\ref{sec:polishresults}).
\item In the future, laptops may become a popular platform for running genome assembly tools, as the portability of a laptop makes it a good fit for in-field analysis. Compared to the desktop and server platforms that we use to test our pipelines, a laptop has even greater memory constraints and lower computational power, and we must factor in limited battery life when evaluating the tools. Based on the scalability studies we perform using our desktop and server platforms, we would likely recommend using Minimap followed by Miniasm for the assembly step, and Minimap followed by Racon for the polishing step, when performing assembly on a laptop. These three tools use relatively low amounts of memory, and execute quickly, which we expect would make the tools a good fit for the various constraints of a laptop. Despite their low memory usage and fast execution, our recommended pipeline can produce high-quality assemblies that are suitable for fast initial in-field analyses. We leave it to future work to quantitatively study the genome assembly pipeline using nanopore sequence data on laptops and other mobile devices.
\end{itemize}
\vspace{-10pt}
\subsection*{Recommendations for Tool Developers}
Based on our analyses, we make the following recommendations for the tool developers.
\vspace{-5pt}
\begin{itemize}
\item The choice of language to implement the tool plays a crucial role regarding the overall performance of the tool. For example, although the basecallers Scrappie and Nanonet belong to the same family (\ie they both use the more accurate RNNs for basecalling), Scrappie is significantly faster than Nanonet since Scrappie is implemented in C whereas Nanonet is implemented in Python (See analysis in Section~\ref{sec:basecallresults}).
\item Memory usage is an important factor that greatly affects the performance and the usability of the tool. While developing new tools or improving the current ones, the developers should be aware of the memory hierarchy. Data structure choices that can minimize the memory requirements and cache-efficient algorithms have a positive impact on the overall performance of the tools. Keeping memory usage in check with the number of threads can enable not only a usable (\ie runnable on machines with relatively small memories) tool but also a fast one. For example, we find that GraphMap cannot even run with a single-thread in our \emph{desktop} machine due to excessively high memory usage (See analyses in Sections~\ref{sec:basecallresults}--~\ref{sec:polishresults}).
\item Scalability of the tool with the number of cores/threads is an important requirement. It is important to make the tool efficiently parallelized to decrease the overall runtime. Design choices should be made wisely while considering the possible overheads that parallelization can add. For example, we find that the parallel speedup of Minimap reduces when the number of threads reaches a high number due to a large increase in the overhead of synchronization between threads (See analyses in Sections~\ref{sec:basecallresults}--~\ref{sec:polishresults}).
\item Since parallelizing the tool can increase the memory usage, dividing the input data into batches, or limiting the memory usage of each thread, or dividing the computation instead of dividing the dataset between simultaneous threads can prevent large increases in memory usage, while providing performance benefits from parallelization. For example, in Nanonet, all of the threads share the computation of each read, and thus memory usage is not affected by the amount of thread parallelism. As a result, Nanonet's usability is not limited to machines with relatively larger memories (See analyses in Sections~\ref{sec:basecallresults}--~\ref{sec:polishresults}).
\end{itemize}

\vspace{-15pt}
\section{Conclusion} \label{sec:conclusion}
We analyze the multiple steps and the associated state-of-the-art tools in the genome assembly pipeline using nanopore sequence data\footnote{We leave it to future work to quantitatively study tools for different applications of nanopore sequencing, such as variant calling, detection of base modifications (\ie methylation studies \cite{simpson2017detecting}), and pathogen detection.} in terms of accuracy, speed, memory efficiency and scalability. We make four major conclusions based on our experimental analyses of the whole pipeline. First, the basecalling tools with higher accuracy and performance, like Scrappie, can overcome the major drawback of nanopore sequencing technology, \ie high error rates. Second, the read-to-read overlap finding tools, Minimap and GraphMap, perform similarly in terms of accuracy. However, Minimap performs better than GraphMap in terms of speed and memory usage by storing only minimizers instead of all \textit{k}-mers, and GraphMap is not scalable when running on machines with relatively small memories. Third, the fast but less accurate assembler Miniasm can be used for a very fast initial assembly, and further polishing can be applied on top of it to increase the accuracy of the final assembly. Fourth, a state-of-the-art polishing tool, Racon, generates high-quality consensus sequences while providing a significant speedup over another polishing tool, Nanopolish.\par
We hope and believe that our observations and analyses will guide researchers and practitioners to make conscious and effective choices while deciding between different tools for each step of the genome assembly pipeline using nanopore sequence data. We also hope that the bottlenecks or the effects of design choices we have found and exposed can help developers in building new tools or improving the current ones.
\vspace{-15pt}
\section*{Key Points} \label{sec:keypoints}
To our knowledge, this is the first work that analyzes state-of-the-art tools associated with each step of the genome assembly pipeline using sequence data generated with nanopore sequencing, a promising new sequencing technology.\par
The key contributions are:\par
1. We analyze the tools in multiple dimensions that are important for both developers and users/practitioners: accuracy, performance, memory usage and scalability.\par
2. We reveal new bottlenecks and tradeoffs that different combinations of tools lead to, based on our extensive experimental analyses.\par
3. We provide guidelines for both practitioners, such that they can determine the appropriate tools and tool combinations that can satisfy their goals, and tool developers, such that they can make design choices to improve current and future tools.\par
4. We show that tools that are aware of the memory hierarchy have a better overall performance and scalability, and they are more usable than the tools that do not keep memory usage in check with the number of threads.\par
5. We show that basecalling is the most important step of the pipeline to overcome the high error rates of nanopore sequencing technology.\par
6. We show that there is a tradeoff between accuracy and performance when choosing the tool for the assembly step. Miniasm, coupled with an additional polishing step can lead to faster overall assembly than using Canu itself, while producing high-quality assemblies.\par

\vspace{-15pt}
\section*{Acknowledgments} \label{sec:acknowledgements}
We thank Jared Simpson and David Matei for their feedback and help with the questions about the tools. Posters describing earlier stages of the work in this paper were presented at PSB 2017 and ISMB-ECCB 2017. We thank the poster session attendees for their feedback on the works. We especially thank Adam M. Phillippy and Mile {\v{S}}iki{\'c} for their feedback during the poster sessions. We also thank developers of Nanonet and Racon for answering our questions on GitHub.

\vspace{-15pt}
\section*{Funding} \label{sec:funding}
This work was supported by a grant from the National Institutes of Health to O.M. and C.A. (HG006004); an installation grant from the European Molecular Biology Organization to C.A. (EMBO-IG 2521); and gifts from Google, Intel, Samsung, and VMware.

\vspace{-15pt}
\begin{footnotesize}
\section*{Author Details}\label{sec:authordetails}
\textbf{Damla Senol Cali} is a PhD student in the Department of Electrical and Computer Engineering at Carnegie Mellon University. Her research interests are in computational methods for the analysis of NGS and nanopore sequencing data, and computer architecture.\par
\textbf{Jeremie S. Kim} is a PhD student in the Department of Electrical and Computer Engineering at Carnegie Mellon University and in the Department of Computer Science at ETH Z\"urich. His research interests are in computer architecture and hardware accelerators for bioinformatics applications.\par
\textbf{Saugata Ghose, PhD,} is a Systems Scientist in the Department of Electrical and Computer Engineering at Carnegie Mellon University. His research interests are in several aspects of computer architecture, with a significant focus on designing architecture-aware and systems-aware memory and storage.\par
\textbf{Can Alkan, PhD,} is an Assistant Professor in the Department of Computer Engineering at Bilkent University. His research interests are in combinatorial algorithms for bioinformatics and computational biology.\par
\textbf{Onur Mutlu, PhD,} is a Professor in the Department of Computer Science at ETH Z\"urich. He is also an Adjunct Professor in the Department of Electrical and Computer Engineering at Carnegie Mellon University. His research interests are in computer architecture, systems, security and bioinformatics.

\end{footnotesize}

\begin{footnotesize}
\bibliographystyle{unsrt}

\bibliography{references}
\end{footnotesize}

\end{document}